\documentclass[preprint,5p,times,twocolumn]{elsarticle}

\usepackage{graphicx}
\graphicspath{{./img/}}
\DeclareGraphicsExtensions{.pdf}
\usepackage{amsmath}
\usepackage{algorithmic}
\usepackage{url}
\usepackage{nohyperref}
\usepackage[group-separator={,},per-mode=symbol]{siunitx}
\usepackage[nohyperlinks,nolist]{acronym}
\usepackage{tikz}
\usetikzlibrary{patterns}
\usetikzlibrary{spy}
\usepackage{collcell}
\usepackage{hhline}
\usepackage{tabularx}
\usepackage{caption}
\newcommand*{\MinNumber}{0.0}%
\newcommand*{\MidNumber}{0.5} %
\newcommand*{\MaxNumber}{1.0}%
\newcommand{\ApplyGradient}[1]{%
        \ifdim #1 pt > \MidNumber pt
            \pgfmathsetmacro{\PercentColor}{max(min(100.0*(#1 - \MidNumber)/(\MaxNumber-\MidNumber),100.0),0.00)} %
            \hspace{-0.33em}\colorbox{green!\PercentColor!yellow}{#1}
        \else
            \pgfmathsetmacro{\PercentColor}{max(min(100.0*(\MidNumber - #1)/(\MidNumber-\MinNumber),100.0),0.00)} %
            \hspace{-0.33em}\colorbox{red!\PercentColor!yellow}{#1}
        \fi
}
\newcolumntype{Y}{>{\centering\arraybackslash}X}
\newcolumntype{L}{>{\arraybackslash}X}
\newcolumntype{R}{>{\raggedleft\arraybackslash}X}
\newcolumntype{C}[1]{>{\centering\arraybackslash}p{#1}}
\newcolumntype{G}[1]{>{\collectcell\ApplyGradient}#1<{\endcollectcell}}

\usepackage{multirow}
\usepackage{booktabs}
\usepackage{pgfplots}
\pgfplotsset{compat=newest}
\usepgfplotslibrary{units}
\makeatletter
\pgfplotsset{
 unit code/.code 2 args=
   \begingroup
   \protected@edef\x{\endgroup\si{#2}}\x
}
\makeatother
\usetikzlibrary{pgfplots.groupplots}
\usetikzlibrary{matrix}
\usepackage[absolute,showboxes]{textpos}

\definecolor{corered}{RGB}{189, 40, 22}
\definecolor{coregray}{RGB}{191, 191, 191}
\definecolor{coredarkgray}{RGB}{51,51,51}
\definecolor{coreblue}{RGB}{1, 49, 121}
\makeatletter
\pgfplotsset{
    groupplot xlabel/.initial={},
    every groupplot x label/.style={
        at={($({group c1r\pgfplots@group@rows.west}|-{group c1r\pgfplots@group@rows.outer south})!0.5!({group c\pgfplots@group@columns r\pgfplots@group@rows.east}|-{group c\pgfplots@group@columns r\pgfplots@group@rows.outer south})$)},
        anchor=north,
    },
    groupplot ylabel/.initial={},
    every groupplot y label/.style={
        rotate=90,
        at={($({group c1r1.north}-|{group c1r1.west})!0.5!({group c1r\pgfplots@group@rows.south}-|{group c1r\pgfplots@group@rows.outer west})$)},
        anchor=south
    },
    execute at end groupplot/.code={%
      \node [/pgfplots/every groupplot x label]
{\pgfkeysvalueof{/pgfplots/groupplot xlabel}};  
      \node [/pgfplots/every groupplot y label] 
{\pgfkeysvalueof{/pgfplots/groupplot ylabel}};  
    },
    group/only outer labels/.style =
{
group/every plot/.code = {%
    \ifnum\pgfplots@group@current@row=\pgfplots@group@rows\else%
        \pgfkeys{xticklabels = {}, xlabel = {}}\fi%
    \ifnum\pgfplots@group@current@column=1\else%
        \pgfkeys{yticklabels = {}, ylabel = {}}\fi%
}
}
}

\def\endpgfplots@environment@groupplot{%
    \endpgfplots@environment@opt%
    \pgfkeys{/pgfplots/execute at end groupplot}%
    \endgroup%
}
\makeatother

\usepackage{pifont}

\usepackage{xspace}
\newcommand{\etal}{\textit{et~al.}~}
\newcommand{\eg}{\textit{e.g.,}~}
\newcommand{\ie}{\textit{i.e.,}~}
\newcommand{\cf}{\textit{cf.,}~}

\newcommand{\one}{({\em i})\xspace}
\newcommand{\two}{({\em ii})\xspace}
\newcommand{\three}{({\em iii})\xspace}
\newcommand{\four}{({\em iv})\xspace}

\newcommand{\documentTitle}
{%
Network Anomaly Detection in Cars:\\A Case for Time-Sensitive Stream Filtering and Policing
}%

\journal{Computer Networks}

\begin{document}

\begin{frontmatter}

\title{
    \documentTitle
}

\setlength{\TPHorizModule}{\paperwidth}
\setlength{\TPVertModule}{\paperheight}
\TPMargin{5pt}
\begin{textblock}{0.8}(0.1,0.02)
     \noindent
     \footnotesize
     If you cite this paper, please use the original reference:
     Philipp Meyer, Timo H\"ackel, Sandra Reider, Franz Korf, and Thomas C. Schmidt. "Network Anomaly Detection in Cars: A Case for Time-Sensitive Stream Filtering and Policing," \emph{Computer Networks}, vol. 255, p. 110855, Dec. 2024, doi: https://doi.org/10.1016/j.comnet.2024.110855.
\end{textblock}

\author[haw]{Philipp~Meyer}
\ead{philipp.meyer@haw-hamburg.de}

\author[haw]{Timo Häckel}
\ead{timo.haeckel@haw-hamburg.de}

\author[haw]{Sandra Reider}
\ead{sandra.reider@haw-hamburg.de}

\author[haw]{Franz Korf}
\ead{franz.korf@haw-hamburg.de}

\author[haw]{Thomas C. Schmidt}
\ead{t.schmidt@haw-hamburg.de}

\affiliation[haw]{organization={Department of Computer Science, Hamburg University of Applied Sciences},%
            addressline={Berliner Tor 7}, 
            city={Hamburg},
            postcode={20099}, 
            state={Hamburg},
            country={Germany}}

\begin{abstract}
Connected vehicles are threatened by  cyber-attacks as in-vehicle networks technologically approach (mobile) LANs with several wireless interconnects to the outside world.
Malware that infiltrates a car today faces potential victims of constrained, barely shielded Electronic Control Units (ECUs).
Many ECUs perform critical driving functions, which stresses the need for hardening security and resilience of in-vehicle networks in a multifaceted way.
Future vehicles will comprise Ethernet backbones that differentiate services via  Time-Sensitive Networking (TSN). 
The well-known vehicular control flows will follow predefined schedules and TSN traffic classifications.	
In this paper, we exploit this traffic classification to build a network anomaly detection system.
We show how filters and policies of TSN can identify misbehaving traffic and thereby serve as distributed guards on the data link layer. 
On this lowest possible layer, our approach  derives a highly efficient network protection directly from TSN.
We classify link layer anomalies and micro-benchmark the detection accuracy in each class.
Based on a topology derived from a real-world car and its traffic definitions we evaluate the detection system in realistic macro-benchmarks based on recorded attack traces.
Our results show that the detection accuracy depends on how exact the specifications of in-vehicle communication are configured. Most notably for a fully specified communication matrix, our anomaly detection  remains free of false-positive alarms, which is a significant benefit for implementing automated countermeasures in future vehicles. 
\end{abstract}


\end{frontmatter}

%%%% 	Acronyms    %%%%
% !TEX root = ../main.tex
\begin{acronym}
	% A
	\acro{ACDC}[ACDC]{Automotive Cyber Defense Center}
	\acro{AD}[AD]{Anomaly Detection}
	\acro{ADS}[ADS]{Anomaly Detection System}
	\acroplural{ADS}[ADSs]{Anomaly Detection Systems}
	\acro{API}[API]{Application Programming Interface}
	\acro{AUTOSAR}[AUTOSAR]{AUTomotive Open System ARchitecture}
	\acro{AVB}[AVB]{Audio Video Bridging}
	\acro{ARP}[ARP]{Address Resolution Protocol}
	% B
	\acro{BE}[BE]{Best-Effort}
	% C
	\acro{CAN}[CAN]{Controller Area Network}
	\acro{CBM}[CBM]{Credit Based Metering}
	\acro{CBS}[CBS]{Credit Based Shaping}
	\acro{CMI}[CMI]{Class Measurement Interval}
	\acro{CoRE}[CoRE]{Communication over Realtime Ethernet}
	\acro{CT}[CT]{Cross Traffic}
	% D
	\acro{DoS}[DoS]{Denial of Service}
	\acro{DDoS}[DDoS]{Distributed \acl{DoS}}
	\acro{DPI}[DPI]{Deep Packet Inspection}
	% E
	\acro{ECU}[ECU]{Electronic Control Unit}
	\acroplural{ECU}[ECUs]{Electronic Control Units}
	% F
	\acro{FN}[FN]{False Negatives}
	\acro{FP}[FP]{False Positives}
	% H
	\acro{HTTP}[HTTP]{Hypertext Transfer Protocol}
	\acro{HMI}[HMI]{Human-Machine Interface}
	% I
	\acro{IA}[IA]{Industrial Automation}
	\acro{IAM}[IAM]{Identity- and Access Management}
	\acro{ICT}[ICT]{Information and Communication Technology}
	\acro{IDS}[IDS]{Intrusion Detection System}
	\acroplural{IDS}[IDSs]{Intrusion Detection Systems}
	\acro{IEEE}[IEEE]{Institute of Electrical and Electronics Engineers}
	\acro{IoT}[IoT]{Internet of Things}
	\acro{IP}[IP]{Internet Protocol}
	\acro{IVN}[IVN]{In-Vehicle Network}
	\acroplural{IVN}[IVNs]{In-Vehicle Networks}
	% J
	% L
	\acro{LIN}[LIN]{Local Interconnect Network}
	% M
	\acro{MTU}[MTU]{maximum transmission unit}
	\acro{MOST}[MOST]{Media Oriented System Transport}
	% N
	\acro{NAD}[NAD]{Network Anomaly Detection}
	\acro{NADS}[NADS]{Network Anomaly Detection System}
	\acroplural{NADS}[NADSs]{Network Anomaly Detection Systems}
	% O
	\acro{OEM}[OEM]{Original Equipment Manufacturer}
	% P
	\acro{PSFP}[PSFP]{per-stream filtering and policing}
	% R
	\acro{RC}[RC]{Rate-Constrained}
	\acro{REST}[ReST]{Representational State Transfer}
	% S
	\acro{SDN}[SDN]{Software-Defined Networking}
	\acro{SOA}[SOA]{Service-Oriented Architecture}
	\acroplural{SOA}[SOAs]{Service-Oriented Architectures}
	\acro{SOME/IP}[SOME/IP]{Scalable service-Oriented MiddlewarE over IP}
	\acro{SR}[SR]{Stream Reservation}
	\acro{SRP}[SRP]{Stream Reservation Protocol}
	% T
	\acro{TAS}[TAS]{Time-Aware Shaping}
	\acro{TCP}[TCP]{Transmission Control Protocol}
	\acro{TDMA}[TDMA]{Time Division Multiple Access}
	\acro{TN}[TN]{True Negatives}
	\acro{TP}[TP]{True Positives}
	\acro{TSN}[TSN]{Time-Sensitive Networking}
	\acro{TSSDN}[TSSDN]{Time-Sensitive Software-Defined Networking}
	\acro{TT}[TT]{Time-Triggered}
	\acro{TTE}[TTE]{Time-Triggered Ethernet}
	% U
	\acro{UDP}[UDP]{User Datagram Protocol}
	% Q
	\acro{QoS}[QoS]{Quality-of-Service}
	% V
	\acro{V2X}[V2X]{Vehicle-to-X}
	% W
	\acro{WS}[WS]{Web Services}
	% Z
	\acro{ZC}[ZC]{Zone-Controller}
	\acroplural{ZC}[ZCs]{Zone-Controllers}
\end{acronym}

%%%%	Document    %%%%
\section{Introduction}\label{sec:introduction} 
Internal networks of current vehicles are based on \ac{CAN} buses, which are assigned to individual domains and interconnect via a central gateway. New functions of modern vehicles --- including advanced driver assistance and autonomous driving (level 3 to 5) --- rely on data streams from cameras, LIDARs, etc.
These applications demand for significantly higher network capacities  and at the same time require massive, time-critical cross-domain communication.

Ethernet is emerging as the core communication technology 
to serve these needs of future \acp{IVN}: Ethernet is scalable, of low complexity, and cheap.
A new, vehicle-specific physical layer  enables transmission speeds of up to \SI{1}{\giga\bit\per\second} via a single twisted pair while also meeting the requirements of electromagnetic compatibility~\cite{ieee8023bw-15,ieee8023bp-16}.
The IEEE \ac{TSN}~\cite{ieee8021q-18}  umbrella standard is a leading candidate to provide robust \ac{QoS} guarantees in \acp{IVN}. 
It is expected that \acp{IVN} will gradually transition to flat Ethernet-centric communication~\cite{mk-ae-15}, after first replacing the central \ac{CAN} gateway by a \ac{TSN} Ethernet backbone, and later dissolving domains into zones that represent  physical locations in the car (\eg front/left/right/rear)~\cite{brkw-aeajr-17}.
In this topology, each zone connects via a \ac{ZC}, which acts as a gateway between the peripheral bus and the Ethernet backbone.
\ac{CAN} messages destined for a different domain in the same zone traverse their single \ac{ZC}.
\ac{CAN} messages destined for another zone, even if the domain is identical, get forwarded via the Ethernet backbone. 

The attack surface of future cars will span various interfaces ranging from near-field communication to the global Internet~\cite{cmkas-ceaas-11}. Advanced, (semi-)autonomous driving heavily depends on data exchange with external peers such as \ac{V2X} or cloud services~\cite{bddk-rcsJR-18}.
For a holistic protection, security mechanisms in different domains and layers are under investigation~\cite{ieee8021ae-18,dh-ijr-02,RFC-8446,m-saejr-17,rgks-oasjr-20,psz-hcdjr-17,so-sagjr-14,lss-eacdc-19,ymhs-tbijr-19}. With the release of the ISO/SAE 21434~\cite{iso-sae-21434} standard for cybersecurity in road vehicles and the European Cybersecurity Act \cite{eu-ecaJR-19}, the industry is compelled to harden security protection of future cars.
In the event of a successful intrusion (via some vehicle interface), monitoring of the in-car communication still can detect  misbehavior and execute countermeasures.
Noteworthily, guarding the internal link layer not only protects the car from external attacks but may also shield against compromised components within the vehicle.

In this paper, we propose and evaluate link layer traffic monitoring based on \ac{TSN} stream classification. 
\ac{TSN} schedules provide a precise notion of critical communication patterns, which we leverage to develop a fine granular view on regular packet flows on the link layer.
Monitoring the link layer is crucial for detecting misbehavior that threatens the safety and security of the vehicle. 
Protecting link layer communication promises the highest efficiency since performed on the lowest available layer---including the robustness of the network performance (\eg \ac{QoS}~\cite{ls-pitjr-19,fk-tsnjr-17,slksh-bhcan-15}) and of all upper layers.
Early work~\cite{mhks-nadci-20} could already demonstrate how \ac{TSN}-based anomaly detection can identify bogus traffic  without false positives.
In this paper, we comprehensively explore the case of a \ac{NADS} based on TSN making the following key contributions.
% \clearpage
\begin{itemize}
    \item We leverage the \ac{TSN} ingress  \ac{PSFP}~\cite{ieee8021q-18,ieee8021qci-17} as the core component of an in-car \ac{NADS}.
    \ac{PSFP} forces all inbound traffic to match regular patterns (\eg timings, bandwidths, packet sizes) by dropping misbehaving frames.
		The \ac{NADS} is established by combining \ac{PSFP} with a central (TSN or SDN) controller.
    \item We micro-benchmark the performance of the \ac{NADS} in simulations.
    To this end, we classify link layer anomalies caused by attacks and individually analyze each class.
    Our results confirm that all detection schemes remain free of false positives. 
    We also show  how false negatives can be reduced by correlating multiple \ac{PSFP} indicators. 
    \item We present a realistic macro-benchmark in a simulated zonal \ac{IVN} topology using attack traces from CIC-IDS~2017~\cite{shg-gniJR-18} and traffic stimuli from the reference communication of an OEM car.
    Our NADS shows promising results throughout the comprehensive simulations, while not a single false alarm occurs. 
\end{itemize}

The remainder of this paper is structured as follows.
In Section \ref{sec:concept_v2}, we introduce the concept of \acf{PSFP} and its potential to detect anomalies.
We classify link layer anomalies in Section \ref{sec:benchmark_v2} and assess the effectiveness of detecting them in differentiated micro-benchmarks. 
In Section \ref{sec:evaluation}, we use an OEM car communication model to  macro-benchmark the behavior of our \ac{NADS} in realistic scenarios.
Section \ref{sec:background_v2_security} discusses the problems of cyber protection for cars along with  related work. 
Finally, we conclude in Section \ref{sec:conclusion} with an outlook on future research directions.

\section{Time Sensitive Networking in Vehicles and its Potential to Detect Anomalies}
\label{sec:concept_v2}

In this section, we introduce the initial idea of how  real-time flow management in  \acp{IVN} can simultaneously identify network anomalies. 
We start from the 
\ac{TSN} standard, which enables \ac{QoS} guarantees in local networks by real-time extensions for IEEE 802.1Q Ethernet~\cite{ieee8021q-18}. 
These extensions comprise mechanisms such as traffic shaping~\cite{ieee8021qbv-16}, ingress control~\cite{ieee8021qci-17} and time synchronization~\cite{ieee8021as-20}.
\ac{TSN} paves the way for  Ethernet \acp{IVN} that comply to timing constraints and strictly meet deadlines of safety critical vehicular applications.
 Time division multiplexing with traffic shaping and prioritization allows synchronous and asynchronous traffic streams of diverging timing constraints to  share the same links successfully.
Provisioning real-time traffic definitions makes it possible to bound delay and jitter for each critical traffic stream.
Synchronous \ac{TDMA} traffic can limit delays in the range of microseconds  and jitter within nanoseconds~\cite{msks-eatts-13}.
For this work, \ac{TSN} standard IEEE 802.1Qci \acf{PSFP}~\cite{ieee8021qci-17} is of particular interest.

In the following, we explain how  \one \ac{PSFP} can enforce individual stream behavior in \ac{TSN} infrastructures, \two  \ac{PSFP} can configure a link layer \ac{NADS}, \three general conditions influence the detection quality, and \four to implement our concept in a \ac{SDN}-based Ethernet backbone.

\subsection{Per-Stream Filtering and Policing}
\label{subsec:concept_v2_psfp}

\begin{figure}
    \centering
    \includegraphics[width=\columnwidth, trim= 0.6cm 0.6cm 0.6cm 0.6cm, clip=true]{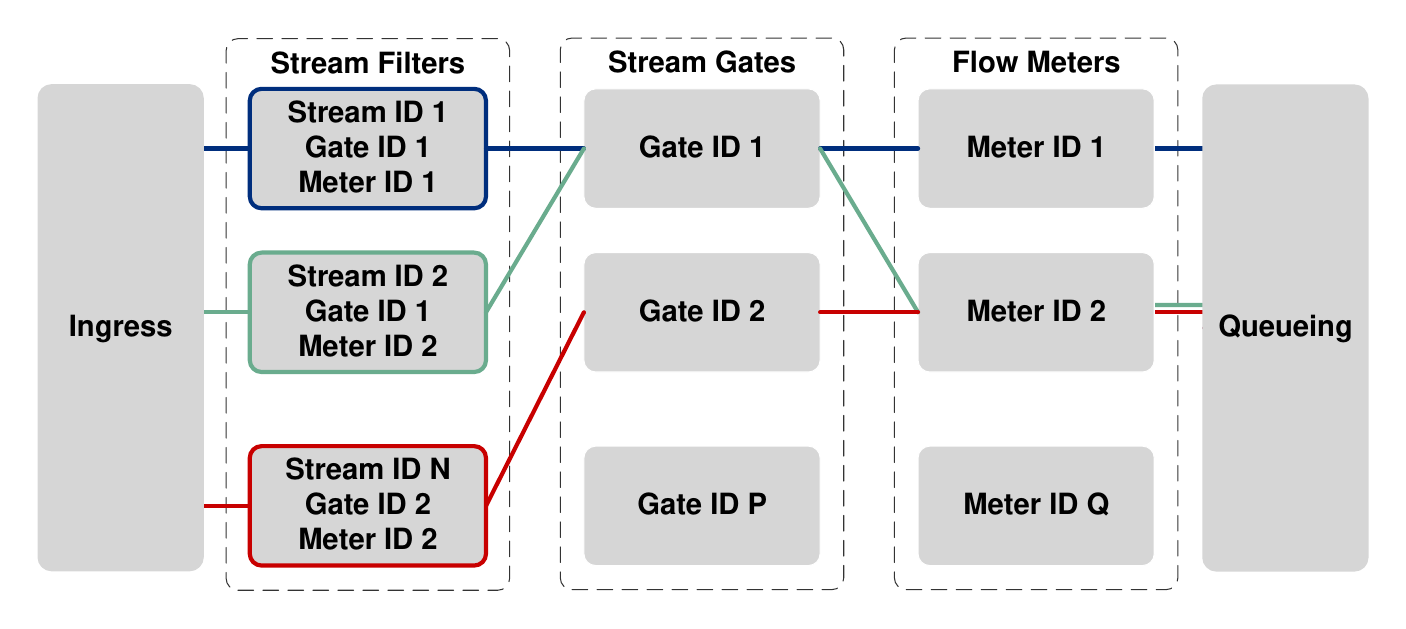}
    \caption{IEEE 802.1Qci \acf{PSFP}}%
    \label{fig:tsn_qci}
\end{figure}

The IEEE 802.1Qci \acf{PSFP} standard~\cite{ieee8021qci-17} controls individual flows  and enforces rules on all incoming traffic. Fig.~\ref{fig:tsn_qci} shows how filtering and policing is applied to ingress traffic in \ac{PSFP}.

A frame needs to pass three stages prior to queueing.
The first stage identifies individual streams and maps them to corresponding gates and meters.
In the second stage, individual gates are responsible for admission of incoming frames.
The state of stream gates can be static or change at runtime depending on a predefined periodic schedule and network-wide synchronized clocks.
If a gate is ``CLOSED'', incoming frames will be dropped.
Once ``OPEN'', a flow meter measures the frames in a third stage.
Flow meters apply algorithms to determine whether a frame is allowed to pass.
For example, a flow meter can limit bandwidth or burst size and 
 would mark or drop a frame that exceeds the bandwidth or burst allowance.

During the \ac{IVN} design process, traffic relations are pre-defined in a communication matrix of the vehicle and annotated with behavioral properties such as timings, bandwidths, and bursts.
The \ac{PSFP} meter and gate configurations must match this traffic specification exactly to assure safety and security.
Exact configurations avoid unwanted disturbance by traffic of misbehaving streams.
Loosely defined configurations could tolerate an over-utilization, which turns harmful for concurrent streams, whereas too strict configurations could lead to dropping frames of valid streams.
We focus on the exact \ac{PSFP} configurations as the basis for in-car network anomaly detection.

\subsection{Anomaly Detection with Filters and Policies}
\label{subsec:concept_v2_concept}

Anomaly detection is a function of an \ac{IDS}, which can monitor from the perspective of a host (H) or a network (N).
There are two flavors of \acp{IDS}~\cite{dzjal-adijr-20}.
\one~Signature detection systems that identify patterns by predefined signatures.
These systems work reliably but are limited to known misbehavior.
\two~\acp{ADS} follow a behavioral approach by using predefined descriptions of regular patterns.
Deviating patterns are then marked as misbehavior, which allows  to also detect zero-day attacks in safety critical systems.
The finer the regular patterns are described, the more accurate can the anomaly detection operate.

Safety-critical communication flows of in-car networks are pre-defined in the network design.
In \ac{PSFP}, each unique traffic flow is represented as a link layer stream.
Hence, the pre-defined link layer behavior of safety-critical communication is already part of the \ac{PSFP} configuration.
From this perspective, the \ac{PSFP} configuration can be read as an implicit description of the behavioral horizon between normal and abnormal link utilization.
To this end, a frame drop by a \ac{PSFP} rule indicates unintended behavior and acts as a detector of an anomaly---a valuable input for an anomaly detector.

False positives in the anomaly detection hinder automated counter measures. They occur whenever traffic classifications need not be rigorously met.
As in-vehicle control traffic is predefined by design, we can 
 focus in the following on traffic characteristics which cars in regular and error-free operations never violate.
Corresponding indicators then operate reliably in the sense that they remain free of false positives, provided the \ac{PSFP} configuration is correct.

Managed switches collect statistics of network events for administrative inspection.
A \ac{TSN} switch also records \ac{TSN}-specific statistics.
The number of individual frame drops of \ac{PSFP} flow meters and stream gates are such values of a \ac{TSN} switch.
Traditional network management can use SNMP~\cite{RFC-3410} or NETCONF~\cite{RFC-6241} to control and monitor local switches.
\ac{SDN}-based networks can also collect data from forwarding devices via OpenFlow~\cite{mabpp-oeicn-08,onfts025-15}.
Network managers or SDN controllers can aggregate statistics of all switches in the network and hence enable an anomaly detection function by applying appropriate algorithms to analyze \ac{PSFP} data.

The central collection of statistics from switches introduces a delay, which adds to the time at which anomalies are first discovered.
In our setting, this delay remains uncritical for the link layer availability since \ac{PSFP} is a self-protecting system that drops frames which violate regular patterns.
This local guarding prevents compromised streams from spreading and negatively impacting concurrent streams.
Since network resource consumption is controlled  by \ac{PSFP}, individual streams cannot exceed predefined resources. 
 This in particular prevents volumetric DoS attacks from spreading in the car network.

Overall, combining \ac{PSFP}  with event monitoring in switches provides a toolbox that can detect anomalies in the network reliably and prevent resource exhaustion. 
It is noteworthy that the underlying base technologies are already in discussion for an \ac{IVN} deployment~\cite{brkw-aeajr-17, hhlng-saeea-20}.
The proposed \acf{NADS} is based on the configuration patterns in \ac{PSFP} for regulating network traffic.

\subsection{Quality of Anomaly Detection}
\label{subsec:concept_v2_potentials}

The strict safety requirements for vehicles on the road challenge the quality of an onboard anomaly detection system.
A \ac{NADS} in the vehicle must achieve high precision while minimizing false alarms.
At the same time, there is no technology that detects all known and unknown attacks.
Hence, a \ac{NADS} must be considered as only one component of a comprehensive security solution.

\ac{PSFP} based network anomaly detection offers the potential to detect a wide range of misbehavior.
Those are not limited to attacks.
Failures, bugs, and configuration errors are also possible sources of deviations from standard behavior.

The \ac{PSFP} configuration quality, \ie correctness and completeness of traffic rules, is the key factor for detection quality.
This configuration, which includes \eg timings, bandwidth, and \ac{MTU} is already required for guaranteed real-time communication and vehicle safety.
During testing phases, a \ac{NADS} can help to expose configuration errors.
After a complete fix of configuration errors, operational network traffic never violates the configured constraints under  normal condition---including worst case scenarios and bursts, \eg in case of edge scenarios like a crash.
A correct and exact \ac{PSFP} configuration that also considers worst case scenarios consequently enables a detection with zero false alarms.

Nevertheless, such an \ac{NADS} is not able to detect every kind of misbehavior.
Bogus traffic, for example, which matches exactly with the specifications of a stream remains undetectable.
We investigate the detection performance and its limits in a micro benchmark for all possible misuses on the link layer, as well as in a macro benchmark based on a realistic in-vehicle network (see Section~\ref{sec:benchmark_v2} and~\ref{sec:evaluation}).

\subsection{Implementation in an SDN Backbone}
\label{subsec:concept_v2_example}

\ac{SDN} enables a programmable control plane and shifts the control logic to a central controller.
In contrast to traditional switches and routers, the central \ac{SDN} controller populates flow tables in forwarding devices using open standards such as OpenFlow~\cite{onfts025-15}.
This drastically increases flexibility and controllability of the network~\cite{hmks-stsdn-22}.
Central knowledge and dynamic forwarding decisions also enable the implementation of countermeasures by reconfiguring the networking in response to alarms.
In recent years, use cases of \ac{SDN} extended to areas such as vehicular networks~\cite{hmg-rsarn-18,hhlng-saeea-20,hmks-stsdn-22} and industrial plants~\cite{gkbh-sdfjr-19}. 

\begin{figure}[h!]
    \centering
    \includegraphics[width=\columnwidth, trim= 0.6cm 0.6cm 0.6cm 0.6cm, clip=true]{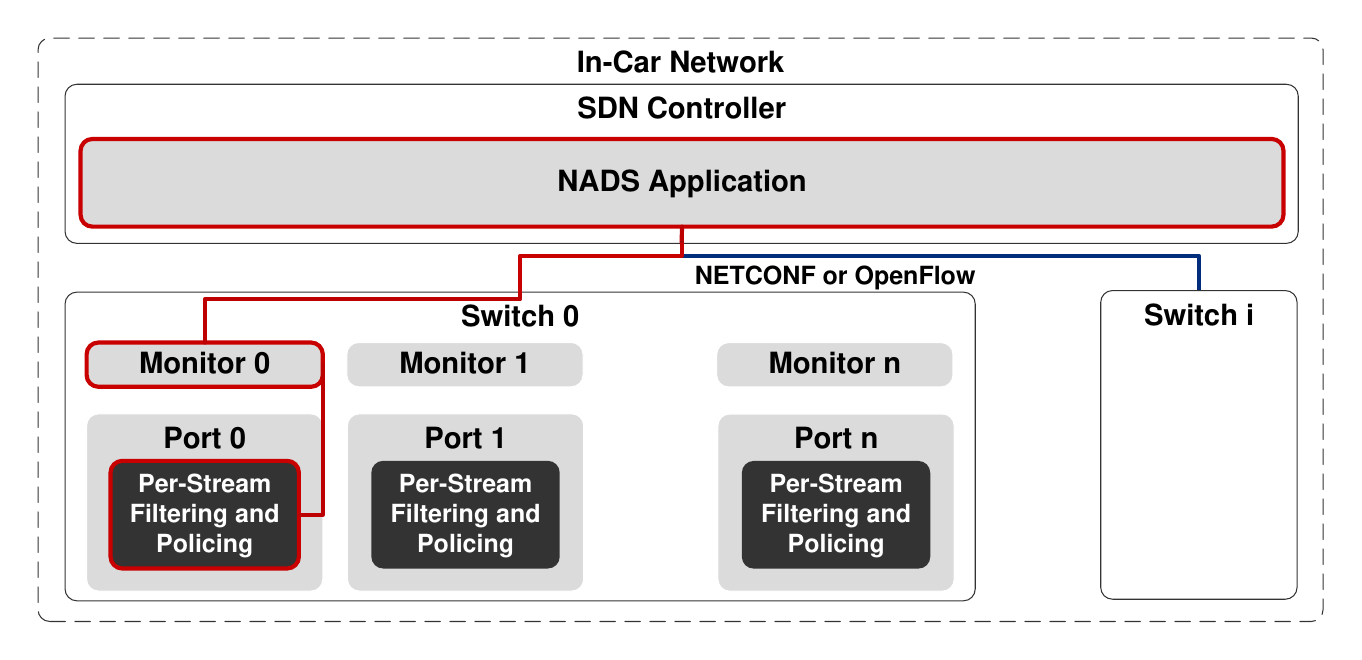}
    \caption{Example of a \ac{PSFP} based \ac{NADS} by combination with \ac{SDN}}%
    \label{fig:qci_sdn}
\end{figure}

In an \ac{SDN}-based in-car backbone, events from \ac{PSFP} instances in individual switches reach the central \ac{SDN} Controller.
For example, an advanced \ac{NADS} application can be deployed at the \ac{SDN} controller without putting additional stress onto the switches or adding additional devices to the network.
Such a \ac{NADS} controller application can use individual events and inspect these in combination with other reports.

Fig.~\ref{fig:qci_sdn} visualizes the operation of a \ac{NADS} with \ac{PSFP} and \ac{SDN} infrastructure.
Each port ($0-n$) of the \ac{TSN} switches $0$ to $i$ has an instance of \ac{PSFP} containing its individual configuration.
All switches monitor events per port of their running \ac{PSFP} instances (\eg frame drops).
The \ac{PSFP} in port $0$ of switch $0$ receives a frame that violates a meter configuration and gets dropped.
This event is monitored and recorded in switch $0$.
Switch $0$ forwards these event statistics upstream to the \ac{SDN} controller.
The \ac{NADS} application of the controller inspects whether the data of the switches indicate a traffic anomaly.
Every increase in the absolute number of dropped frames since network initialization is considered an anomaly.
No additional resources are needed in switches and limited data processing is required in the \ac{SDN} controller to interpret the counters. 
Whenever an anomaly is detected, the \ac{NADS} controller application could send combined reports to a higher instance (\eg a cloud defense center~\cite{lss-eacdc-19}), or initiate countermeasures by reconfiguring the network flow tables or the \ac{TSN} settings.

\section{Micro-Benchmarking Anomaly Detection}
\label{sec:benchmark_v2}

We now proceed to a systematic exploration of the  performance for our network anomaly detectors. We evaluate how reliably \ac{PSFP} statistics can serve as link layer monitors in the presence of attacks.
Root causes of attacks could be malicious \acp{ECU} or external attackers. 
We first characterize the anomaly classes, the assessment metrics, and the simulation environment. Then we present the results of all dedicated benchmarks.

\subsection{Classification of Link Layer Anomalies}
\label{subsec:attacks}
Misbehaving traffic traversing a network influences links and becomes visible at ingress points.
We classify these interactions based on common link layer operations.
All possible operations on Ethernet frames are: Elimination, injection, inspection, manipulation, redirection, reordering, and rescheduling.
The impact of any attack traversing the Ethernet layer falls into at least one of these classes.

\begin{figure}[h!]
    \centering
    \vspace{-5pt}
    \includegraphics[width=\columnwidth, trim= 0.6cm 0.6cm 0.6cm 0.6cm, clip=true]{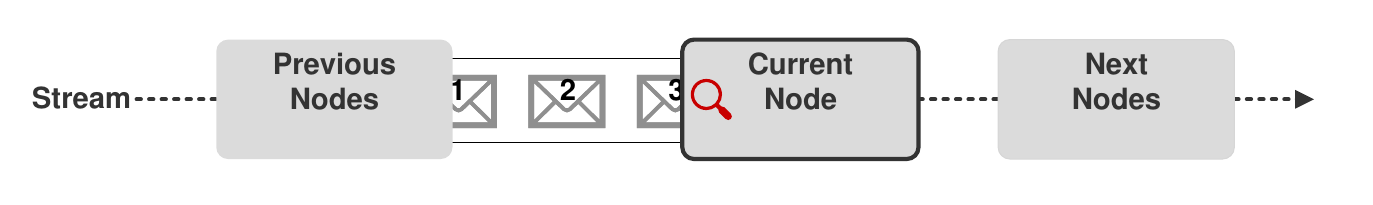}
    \vspace{-15pt}
    \caption{A stream traversing a network without impairment}%
    \label{fig:attack_impact_base}
    \vspace{-15pt}
\end{figure}

Fig.~\ref{fig:attack_impact_base} shows a stream and its frames traversing the network in a baseline scenario without impairment.
Our focus is on an intermediate, ``current'' node, at which  a \ac{TSN} \ac{PSFP} inspects all incoming frames.

\subsubsection{Elimination}
\label{subsubsec:class_elimination}
\begin{figure}[h!]
    \centering
    % \vspace{-5pt}
    \includegraphics[width=\columnwidth, trim= 0.6cm 0.6cm 0.6cm 0.6cm, clip=true]{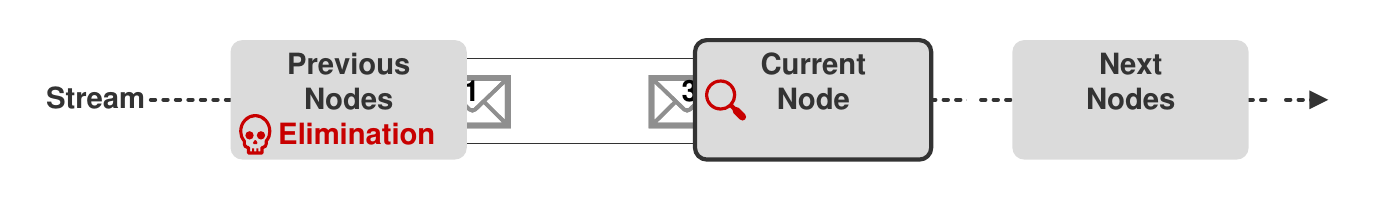}
    \vspace{-15pt}
    \caption{A traversing stream is impaired by eliminated frames}%
    \label{fig:attack_impact_elimination}
    \vspace{-15pt}
\end{figure}

An anomaly that impairs a stream by removing frames.
In  Fig.~\ref{fig:attack_impact_elimination}, frame number 2 is eliminated by one of the previous nodes and detected as missing at the current node. This observation corresponds to blackholing scenarios in wide area networks.

\subsubsection{Injection}
\label{subsubsec:class_injection}
\begin{figure}[h!]
    \centering
    \vspace{-5pt}
    \includegraphics[width=\columnwidth, trim= 0.6cm 0.6cm 0.6cm 0.6cm, clip=true]{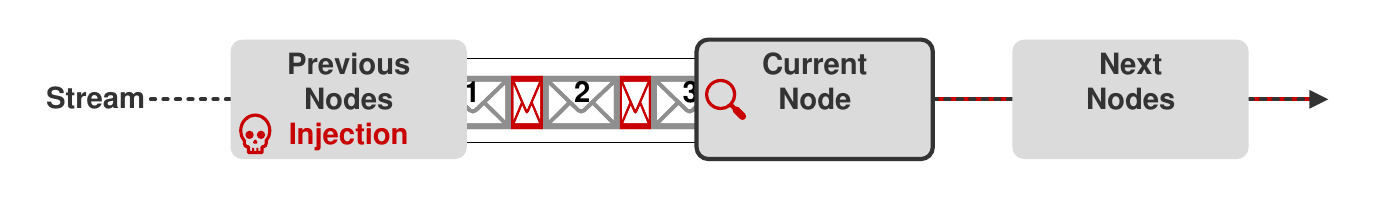}
    \vspace{-15pt}
    \caption{A traversing stream is impaired by injected new frames}%
    \label{fig:attack_impact_injection}
    \vspace{-15pt}
\end{figure}

A stream modification by inserting frames at a previous node.
In  Fig.~\ref{fig:attack_impact_injection}, new frames are injected between frame 1, 2, and 3 as observable by the current node.
Common examples for this class are \ac{DoS} and Distributed \ac{DoS} (\acsu{DDoS}) attacks, but also 
replay attacks.

\subsubsection{Inspection}
\label{subsubsec:class_inspection}
\begin{figure}[h!]
    \centering
    \vspace{-5pt}
    \includegraphics[width=\columnwidth, trim= 0.6cm 0.6cm 0.6cm 0.6cm, clip=true]{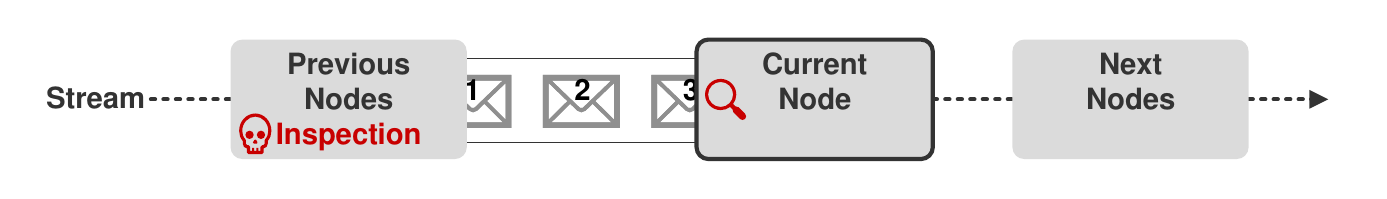}
    \vspace{-15pt}
    \caption{A traversing stream is impaired by an inspection of frames}%
    \label{fig:attack_impact_inspection}
    \vspace{-15pt}
\end{figure}

An unauthorized observation of frames.
The traditional example for this class is eavesdropping.
The actual inspection introduces no changes to the observed stream (see Fig.~\ref{fig:attack_impact_inspection}) but compromises its confidentiality.
The current node is not able to observe this impairment.

\subsubsection{Manipulation}
\label{subsubsec:class_manipulation}
\begin{figure}[h!]
    \centering
    \vspace{-5pt}
    \includegraphics[width=\columnwidth, trim= 0.6cm 0.6cm 0.6cm 0.6cm, clip=true]{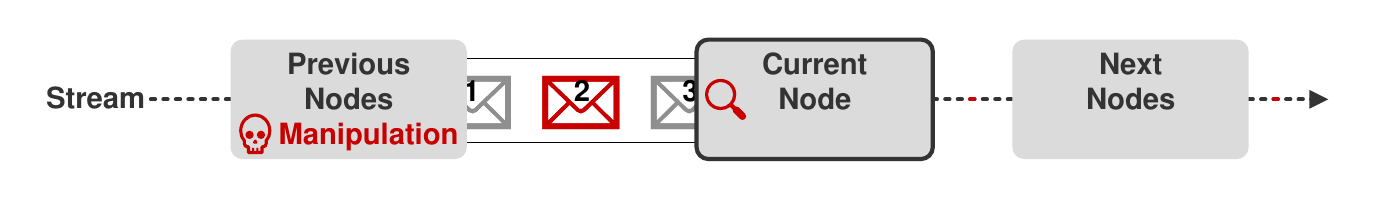}
    \vspace{-15pt}
    \caption{A traversing stream is impaired by manipulation of frames}%
    \label{fig:attack_impact_manipulation}
    \vspace{-15pt}
\end{figure}

A modification of frame payloads.
In Fig.~\ref{fig:attack_impact_manipulation}, a stream is impaired by manipulating  frame 2 at a previous node.
This class includes attacks such as spoofing or application data corruption. 
If corresponding header values (\eg checksum and length) remain unadjusted, or manipulated packets violate the constraints of the stream class, manipulation of frames can be observed at the current node. Manipulations that change the stream affiliation will either appear as a combination of stream elimination and injection, or result in unknown streams.

\subsubsection{Redirection}
\label{subsubsec:class_redirection}
\begin{figure}[h!]
    \centering
    \includegraphics[width=\columnwidth, trim= 0.6cm 1.5cm 0.6cm 0.6cm, clip=true]{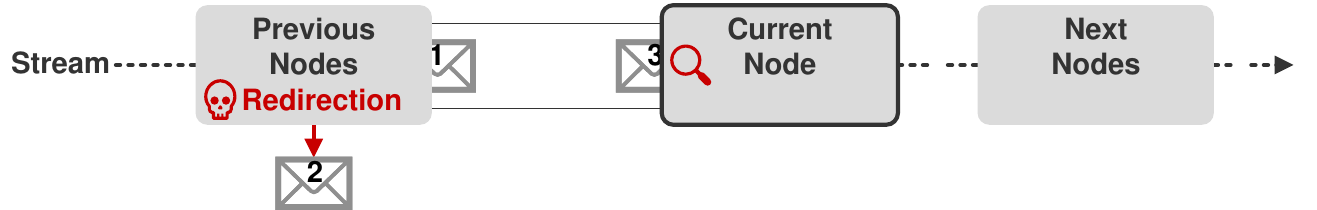}
    \vspace{-15pt}
    \caption{A traversing stream is impaired by frame redirection}%
    \label{fig:attack_impact_redirection}
    \vspace{-15pt}
\end{figure}

A rerouting of frames to a differing path on the network.
In Fig.~\ref{fig:attack_impact_redirection}, frame number 2 is redirected to another next hop by a previous node.
Hence, frame 2 is missing in the stream reaching the current node, which can be observed but remains indistinguishable from frame elimination. On the counter side, redirection appears as an injection.
A wide area network correspondence for this attack is route  poisoning or hijacking.
In a local \ac{SDN} topology a compromised controller can also manipulate the flow tables in the forwarding devices.

\subsubsection{Reordering}
\label{subsubsec:class_reordering}
\begin{figure}[h!]
    \centering
    \includegraphics[width=\columnwidth, trim= 0.6cm 0.6cm 0.6cm 0.6cm, clip=true]{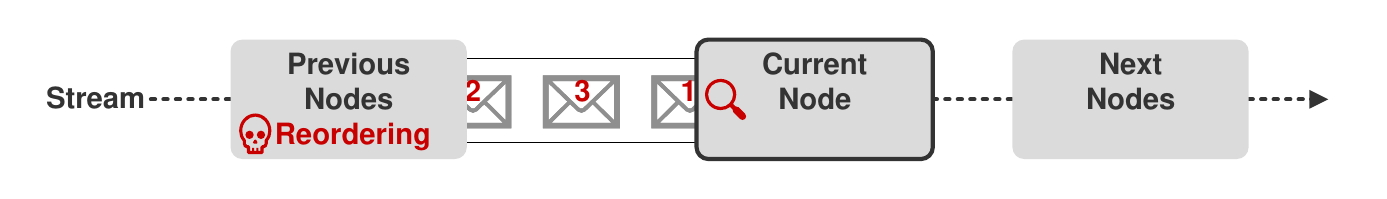}
    \vspace{-15pt}
    \caption{A traversing stream is impaired by frame reordering}%
    \label{fig:attack_impact_reordering}
    \vspace{-15pt}
\end{figure}

A change in sequence of frames.
In Fig.~\ref{fig:attack_impact_reordering}, the frame order is changed from $1\rightarrow 2\rightarrow 3$ to $2\rightarrow 3\rightarrow 1$.
An attack example for reordering is message sequence violation of communication protocols.
If not executed at the source, a frame is taken and inserted after any number of other frames passed.
Reordering of frames is observable at the current node in particular as it changes the timing of frames.

\subsubsection{Rescheduling}
\label{subsubsec:class_timing}
\begin{figure}[h!]
    \centering
    \vspace{-5pt}
    \includegraphics[width=\columnwidth, trim= 0.6cm 0.6cm 0.6cm 0.6cm, clip=true]{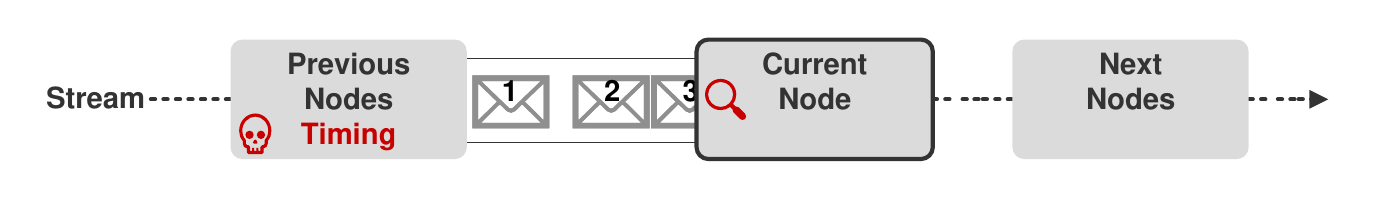}
    \vspace{-15pt}
    \caption{A traversing stream is impaired by timing alteration of frames}%
    \label{fig:attack_impact_timing}
\end{figure}

A change of the forwarding time for individual frames in a stream.
In Fig.~\ref{fig:attack_impact_timing},  frames 1 and 2 are slightly delayed.
This also could be an impact of a malicious manipulation of drift clocks via the time synchronization protocol.
Changes in timing are observable at the current node.

\subsection{Metrics}
\label{subsec:detection_metrics}
To evaluate and compare our classification results we apply common metrics, which are also known from the assessment  for automotive \ac{IDS} performance authored by the U.S. National Highway Traffic Safety Administration~\cite{hgsd-amajr-19}.
Table~\ref{tab:metrics} summarizes these metrics.

\begin{table}[h]
    \caption{Metrics Summary}
    \label{tab:metrics}
    \centering
    \begin{tabularx}{\columnwidth}{lcL}
    \toprule
    \textbf{Metric} & & \textbf{Description}\\
    \midrule
    False Negatives & (FN) & No. of unidentified misbehavior\\
    False Positives & (FP) & No. of false alarms\\ 
    True Positives & (TP) & No. of identified misbehavior\\
    Precision & & Fraction of alarms that identified misbehavior\\
    Recall & & Fraction of identified out of all misbehavior\\
    \bottomrule
    \end{tabularx}
\end{table}

Precision represents the accuracy of identified alarms.
Without \ac{FP} the precision equals 1, but decreases with increasing \acp{FP}:
\begin{equation*}
    Precision = \frac{TP}{TP + FP}
\end{equation*}

Recall represents the fraction of detected attacks from all attacks.
Without  \ac{FN} the Recall equals 1, but decreases with increasing \acp{FN}:
\begin{equation*}
    Recall = \frac{TP}{TP + FN}
\end{equation*}

\paragraph{In-Vehicle Security Implications}
In driving operations,  \acp{FP} in combination with automated countermeasures can cause damage.
Even a very small \ac{FP} rate like 0,001\% for a message with a viable cycle of \SI{100}{\milli\second} leads to an average false positive every 2.8 operating hours.
This may quickly degrade confidence in the \ac{AD} mechanism.

Due to the strict safety requirements for vehicles, it is important to establish a highly trustworthy anomaly detection system.
In particular, a \ac{NADS} in the vehicle must achieve high precision.
As of today, there is no technology that can detect all known and unknown misbehavior in a system, which -- outside of test environments -- always leads to realistic recall values $<1$.
From this follows that it is impossible to operate with zero \acp{FN}.
Anomaly detection systems try to maximize detection rates by their ability to detect zero-day attacks.

Every \ac{NADS} is limited, and \acp{FN} could allow a malicious attacker or device malfunction to cause harm without detecting it.
It is therefore even more important to measure the limits of these systems to be able to determine the risk of unidentified misbehavior.
To protect future cars, a \ac{NADS} is only one component in a comprehensive security solution consisting of various systems and methods. 

\subsection{Simulation Environment and its Configuration}
\label{subsec:detection_environment}
\begin{figure}[h]
    \centering
    \includegraphics[width=\columnwidth, trim=0.6cm 0.6cm 0.6cm 0.6cm, clip=true]{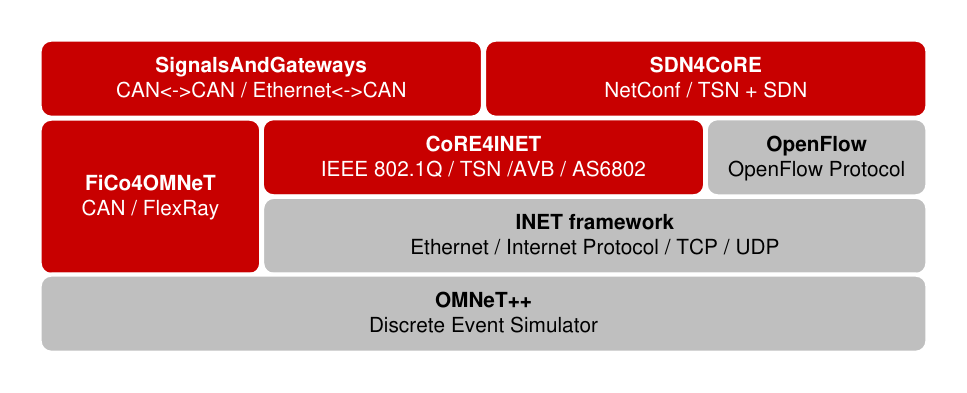}
    \caption{Simulation environment for micro- and macro-benchmarks}%
    \label{fig:simulation_frameworks}
\end{figure}

We use the discrete event simulator OMNeT++ for the execution of all benchmarks. 
Fig.~\ref{fig:simulation_frameworks} shows the frameworks in use for the simulations.
We use this environment as the base for all simulations in this paper.

Basis of the environment is the OMNeT++ discrete event simulator. The combination with the INET framework~\cite{inet-framework} enables the simulation of Ethernet networks.
In the past years, we have built and published a comprehensive open-source environment~\footnote{\url{https://github.com/CoRE-RG}} for simulating \acp{IVN} in OMNeT++~\cite{mkss-smcin-19}.
The frameworks marked in red (CoRE4INET, FiCo4OMNeT, SignalsAndGateways, and SDN4CoRE) are parts of this past work.
They enable the concurrent simulation of real-time Ethernet (\eg \ac{TSN}), field busses (\eg \ac{CAN}), gateways (\eg CAN/Ethernet Gateways), and \ac{SDN}.

\begin{figure}[h]
    \centering
    \includegraphics[width=\columnwidth, trim=0.6cm 0.6cm 0.6cm 0.6cm, clip=true]{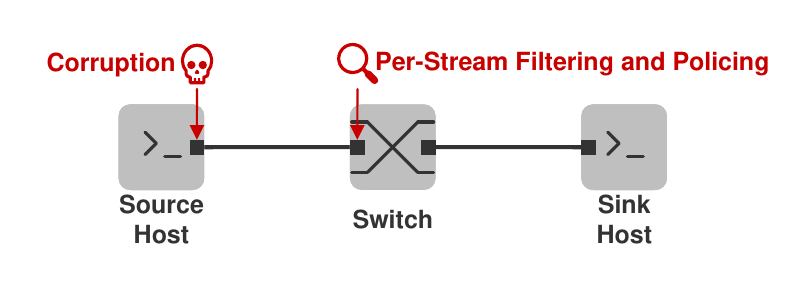}
    \caption{Topology of the simulated network for the micro-benchmark}%
    \label{fig:benchmark_topology}
\end{figure}

In a first step, our goal is to keep the topology as simple as possible for executing micro-benchmarks.
Later in Section~\ref{sec:evaluation}, we assess performance values in macro-benchmarks using a real-world \ac{IVN} topology.
Fig.~\ref{fig:benchmark_topology} depicts the topology for all micro-benchmark simulations.

The initial topology consists of three nodes linked via \SI{100}{\mega\bit\per\second} Ethernet.
Since different Ethernet speeds just rescale, the findings will remain applicable for lower and higher maximum Ethernet throughput.
All traffic is generated at the source host and directed through a switch to the sink host.
To assess how the traffic type affects detection performance, three different traffic patterns and background cross-traffic are transmitted from the source host:
\begin{itemize}
    \item \textbf{Timed control traffic:}
    Maximal Ethernet frames forwarded with the highest 802.1Q priority (7) in a synchronized \ac{TDMA} schedule that reserves time slots on the links in a static period of \SI{0.5}{\milli\second}.
    All other traffic shares the remaining time slots according to their priorities.
    \item \textbf{Shaped data stream:}
    A data stream with a bandwidth of $\sim$\SI{17}{\mega\bit\per\second}.
    802.1Q priority is set to 6 and the traffic is shaped by \ac{CBS}.
    The \ac{CBS} maintains a credit value that ensures a stream does not exceed its reserved bandwidth, while still allowing bursts to make up for time slots where the bandwidth was not reached due to concurrent traffic.
    The remaining bandwidth is available to lower priorities.
    \item \textbf{Prioritized \ac{CAN} tunnel:}
    Legacy CAN messages encapsulated in Ethernet frames with an 802.1Q priority set to 5.
    Messages are generated in a period of \SI{0.5}{\milli\second}.
    \item \textbf{Background cross traffic:}
    Cross traffic generates network load and  concurrently increases the mix of traffic.
    It is generated without 802.1Q tag from uniformly distributed packet sizes and intervals (\SI{125}{\micro\second}--\SI{500}{\micro\second}).
\end{itemize}
The switch operates with pre-configured \ac{PSFP} at the port that connects the source host.
The following single configurations per traffic type enforce key properties of a valid behavior:
\begin{itemize}
    \item \textbf{Timed control traffic:}
    The states of all \ac{PSFP} stream gates depend on a synchronized gate control list.
    Every frame that arrives at the switch port during a closed gate period gets dropped.
    Regular \ac{TDMA} traffic always reaches the respective stream gate in  open state.
    \item \textbf{Shaped data stream:}
    As for the \ac{CBS}, a \ac{CBM}~\cite{mhks-dpcbm-19} is applied to meter a \ac{PSFP} flow and enforce bandwidth reservation.
    Frames that  exceed the reserved bandwidth for a stream get dropped.
    \item \textbf{Prioritized \ac{CAN} tunnel:}
    An encapsulated CAN message (maximum \SI{16}{\byte}) always results in a minimum Ethernet frame. 
    Therefore, \ac{PSFP} is configured to drop all frames with a size that exceeds \SI{64}{\byte}.
    \item \textbf{Background cross traffic:}
    Stream filters match to known cross traffic streams and grant passage through \ac{PSFP} without additional metering rules.
    \item \textbf{Undefined stream:}
    \ac{PSFP} drops all traffic that does not match any stream filter.  
\end{itemize}

\begin{table}[h]
    \caption{Overview of Anomalies in our Benchmarks}
    \label{tab:benchmarked_attack_impacts}
    \centering
    \setlength{\tabcolsep}{3pt}
    \begin{tabularx}{\columnwidth}{lp{3.8cm}Y}
    \toprule
    \centering{\textbf{Anomaly}} & \centering{\textbf{Details}} & \textbf{Benchmark}\\
    \midrule
    \ref{subsubsec:class_elimination} Elimination & Deletion of frames & Yes \\
    \ref{subsubsec:class_injection} Injection & Creation of extra frames & Yes \\
    \ref{subsubsec:class_inspection} Inspection & Observing frames and contents: Not observable in stream behavior & No \\
    \ref{subsubsec:class_manipulation} Manipulation & Alteration of frames & Yes \\
    \ref{subsubsec:class_redirection} Redirection & Adjusting route of frames:\newline Indistinguishable from \ref{subsubsec:class_elimination}, \ref{subsubsec:class_injection}, or \ref{subsubsec:class_manipulation} & No \\
    \ref{subsubsec:class_reordering} Reordering & Changing frame sequences & Yes \\
    \ref{subsubsec:class_timing} Rescheduling & Delay frames & Yes \\
    \bottomrule
    \end{tabularx}
\end{table}

Table \ref{tab:benchmarked_attack_impacts} gives an overview of anomalies that will be analyzed in our benchmarks.
Five of the seven mechanisms are implemented in simulation.
We exclude inspection (\cf Section \ref{subsubsec:class_inspection}) and redirection (\cf Section \ref{subsubsec:class_redirection}).
Inspection has no effect on the behavior of traffic streams and hence remains unobservable.
Redirection is excluded because it cannot be distinguished from elimination, injection, or manipulation.

\begin{figure}[h]
    \centering
    \includegraphics[width=\columnwidth, trim=0.6cm 0.6cm 0.6cm 0.6cm, clip=true]{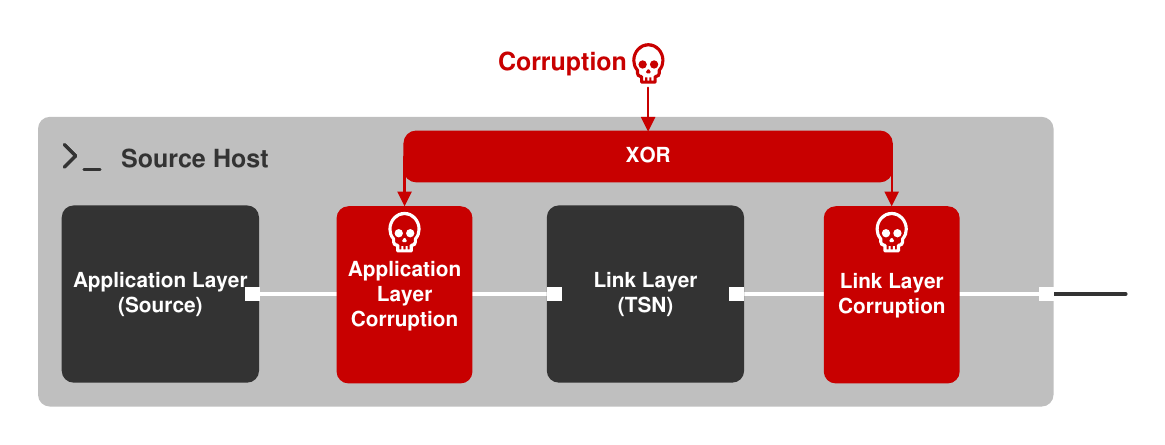}
    \caption{Placement of the corruption layers for simulation in source host}%
    \label{fig:corruption_layer}
\end{figure}

\begin{table*}
    \caption{Drop detection micro-benchmark results with True Positives (TP), False Negatives (FN), and \textbf{Recall (R)}\\
    \textbf{For all simulations:} False Positives $=$ 0 $\Longrightarrow$ \textbf{Precision $=$ \colorbox{green!100}{1.0}}}
    \label{tab:drop_detection_benchmark_results}
    % \vspace{-10pt}
    \begin{tabularx}{\linewidth}{p{3.5cm}YYG{c}YYG{c}YYG{c}YYG{c}YYG{c}}
        \toprule
        \multirow{2}*{\textbf{Traffic pattern}} & \multicolumn{3}{c}{\textbf{Elimination}} & \multicolumn{3}{c}{\textbf{Injection}} & \multicolumn{3}{c}{\textbf{Manipulation}} & \multicolumn{3}{c}{\textbf{Reordering}} & \multicolumn{3}{c}{\textbf{Rescheduling}}\\
        \cmidrule(rl){2-4}\cmidrule(rl){5-7}\cmidrule(rl){8-10}\cmidrule(rl){11-13}\cmidrule(rl){14-16}
        & \multicolumn{1}{c}{TP} & \multicolumn{1}{c}{FN} & \multicolumn{1}{c}{\textbf{R}} & \multicolumn{1}{c}{TP} & \multicolumn{1}{c}{FN} & \multicolumn{1}{c}{\textbf{R}} & \multicolumn{1}{c}{TP} & \multicolumn{1}{c}{FN} & \multicolumn{1}{c}{\textbf{R}} & \multicolumn{1}{c}{TP} & \multicolumn{1}{c}{FN} & \multicolumn{1}{c}{\textbf{R}} & \multicolumn{1}{c}{TP} & \multicolumn{1}{c}{FN} & \multicolumn{1}{c}{\textbf{R}}\\
        \midrule
        \multicolumn{16}{c}{\textbf{Application Layer Corruption}}\\
        \cmidrule(rl){2-4}\cmidrule(rl){5-7}\cmidrule(rl){8-10}\cmidrule(rl){11-13}\cmidrule(rl){14-16}
        \textbf{Timed control traffic} & 0 & 6646 & 0.0 & 6814 & 2602 & 0.72 & 5356 & 1254 & 0.81 & 0 & 6635 & 0.0 & 3 & 6622 & 0.0 \\
        \textbf{Shaped data stream} & 0 & 8918 & 0.0 & 0 & 9412 & 0.0 & 0 & 8880 & 0.0 & 0 & 8875 & 0.0 & 0 & 8882 & 0.0 \\ 
        \textbf{Prioritized CAN tunnel} & 0 & 6664 & 0.0 & 9145 & 263 & 0.97 & 6489 & 173 & 0.97 & 0 & 6659 & 0.0 & 0 & 6657 & 0.0 \\  
        \cmidrule(rl){2-4}\cmidrule(rl){5-7}\cmidrule(rl){8-10}\cmidrule(rl){11-13}\cmidrule(rl){14-16}
        \midrule
        \multicolumn{16}{c}{\textbf{Link Layer Corruption}}\\
        \cmidrule(rl){2-4}\cmidrule(rl){5-7}\cmidrule(rl){8-10}\cmidrule(rl){11-13}\cmidrule(rl){14-16}
        \textbf{Timed control traffic} & 0 & 5656 & 0.0 & 8701 & 706 & 0.92 & 4606 & 1073 & 0.81 & 5616 & 1 & 1.0 & 4382 & 1253 & 0.78 \\
        \textbf{Shaped data stream} & 0 & 8367 & 0.0 & 8931 & 483 & 0.95 & 5862 & 2513 & 0.7 & 2 & 8389 & 0.0 & 1 & 8378 & 0.0 \\
        \textbf{Prioritized CAN tunnel} & 0 & 5616 & 0.0 & 9204 & 207 & 0.98 & 5584 & 165 & 0.97 & 0 & 5696 & 0.0 & 0 & 5692 & 0.0 \\     
        \cmidrule(rl){2-4}\cmidrule(rl){5-7}\cmidrule(rl){8-10}\cmidrule(rl){11-13}\cmidrule(rl){14-16}
        \bottomrule
    \end{tabularx}
\end{table*}

Fig.~\ref{fig:corruption_layer} shows stream corruptions implemented on two different layers within the stack of the source host.
The frames for each stream are generated at the application layer and forwarded onto the link layer before leaving the device.
To cover all elements of compromise, modifications of a stream can be configured to occur between application layer and link layer (application layer corruption), or between link layer and physical port (link layer corruption).
% This way, we cover corruptions before and after the \ac{TSN} link layer.
As the major difference, application layer corruptions need to pass \ac{TSN} traffic shaping, while link layer corruptions are conducted thereafter. In our benchmarks, each case will be executed on both layers.

\subsection{Results}
\label{subsec:detection_results}
The results of the micro-benchmarks are presented first from a baseline scenario without modifications and  from  five scenarios with corruptions thereafter.

The baseline scenario contains only regular traffic, \ie all streams behave as designated.
For this reference benchmark, ten simulation runs (\SI{10}{\second} length) with randomly varying conditions are executed to maximize state coverage.
The results show zero frame drops in all simulations by any module including \ac{PSFP} (\cf Table \ref{tab:benchmarked_regular_traffic}), which implies that the configuration rule set of \ac{PSFP} is consistent with the regular in-car traffic. Accordingly, \ac{FP}=0 for all micro-benchmarks, which implies a perfect precision value of 1:

\begin{equation*}
    Precision = \frac{TP}{TP + FP} = \frac{TP}{TP + 0} = 1
\end{equation*}

It should be noted, though, that legitimate frames may still be dropped in the presence of malicious streams---an incident of misbehavior, which is not a false-positive.  

If the \ac{NADS} detects misbehavior, a precision value of 1 indicates a highly trustworthy system.
To evaluate the overall \ac{NADS} performance, however, the number of false negatives and the recall must also be considered.

In each simulation (\SI{10}{\second} length) with corruption exactly one of the three traffic patterns (timed control traffic, shaped data stream, prioritized CAN tunnel) is modified on one layer.
For five different attacks, this results in thirty different simulation runs.
All stream modifications and frame drops are counted and recorded during the simulations.
The corrupted events are randomized to maximize state coverage.
They take a minimum interval of \SI{1}{\milli\second} to make them distinguishable for analysis.

The numbers of randomly corrupted events range from 5000 to 10000 per simulation.
Only events from \ac{PSFP} are used as anomaly indicators.
In a first step, only frame drop events are considered.
Table \ref{tab:drop_detection_benchmark_results} summarizes the results.

\begin{table}
    \caption{No. of \ac{PSFP} frame drops (FPs) in the baseline benchmark}
    \label{tab:benchmarked_regular_traffic}
    \centering
    \begin{tabularx}{\columnwidth}{lYYY}
    \toprule
    \multirow{2}*{\textbf{Simulation run}} & \multicolumn{3}{c}{\textbf{No. of frames in \ac{PSFP}}}\\
    \cmidrule(rl){2-4}
                                           & \centering{\textbf{Arrived}} & \textbf{Forwarded} & \textbf{Dropped (\ac{FP})}\\
    \midrule
    \#0 --- \#9 & \num{1199688} & \num{1199688} & \num{0} \\
    \bottomrule
    \end{tabularx}
\end{table}

\begin{table*}
    \caption{Drop \& TDMA loss combination detection micro-benchmark results with True Positives (TP), False Negatives (FN), and \textbf{Recall (R)}\\
    \textbf{For all simulations:} False Positives $=$ 0 $\Longrightarrow$ \textbf{Precision $=$ \colorbox{green!100}{1.0}}}
    \label{tab:tdma_combi_detection_benchmark_results}
    % \vspace{-10pt}
    \begin{tabularx}{\linewidth}{p{3.5cm}YYG{c}YYG{c}YYG{c}YYG{c}YYG{c}}
        \toprule
        \multirow{2}*{\textbf{Traffic pattern}} & \multicolumn{3}{c}{\textbf{Elimination}} & \multicolumn{3}{c}{\textbf{Injection}} & \multicolumn{3}{c}{\textbf{Manipulation}} & \multicolumn{3}{c}{\textbf{Reordering}} & \multicolumn{3}{c}{\textbf{Rescheduling}}\\
        \cmidrule(rl){2-4}\cmidrule(rl){5-7}\cmidrule(rl){8-10}\cmidrule(rl){11-13}\cmidrule(rl){14-16}
        & \multicolumn{1}{c}{TP} & \multicolumn{1}{c}{FN} & \multicolumn{1}{c}{\textbf{R}} & \multicolumn{1}{c}{TP} & \multicolumn{1}{c}{FN} & \multicolumn{1}{c}{\textbf{R}} & \multicolumn{1}{c}{TP} & \multicolumn{1}{c}{FN} & \multicolumn{1}{c}{\textbf{R}} & \multicolumn{1}{c}{TP} & \multicolumn{1}{c}{FN} & \multicolumn{1}{c}{\textbf{R}} & \multicolumn{1}{c}{TP} & \multicolumn{1}{c}{FN} & \multicolumn{1}{c}{\textbf{R}}\\
        \midrule
        \multicolumn{16}{c}{\textbf{Application Layer Corruption}}\\
        \cmidrule(rl){2-4}\cmidrule(rl){5-7}\cmidrule(rl){8-10}\cmidrule(rl){11-13}\cmidrule(rl){14-16}
        \textbf{Timed control traffic} & 6645 & 1 & 1.0 & 6818 & 2598 & 0.72 & 5356 & 1254 & 0.81 & 1 & 6634 & 0.0 & 4 & 6621 & 0.0 \\
        \cmidrule(rl){2-4}\cmidrule(rl){5-7}\cmidrule(rl){8-10}\cmidrule(rl){11-13}\cmidrule(rl){14-16}
        \midrule
        \multicolumn{16}{c}{\textbf{Link Layer Corruption}}\\
        \cmidrule(rl){2-4}\cmidrule(rl){5-7}\cmidrule(rl){8-10}\cmidrule(rl){11-13}\cmidrule(rl){14-16}
        \textbf{Timed control traffic} & 5656 & 0 & 1.0 & 8701 & 706 & 0.92 & 4606 & 1073 & 0.81 & 5617 & 0 & 1.0 & 4382 & 1253 & 0.78 \\
        \cmidrule(rl){2-4}\cmidrule(rl){5-7}\cmidrule(rl){8-10}\cmidrule(rl){11-13}\cmidrule(rl){14-16}
        \bottomrule
    \end{tabularx}
\end{table*}

\paragraph{Elimination}
Elimination is configured to delete frames of the observed stream with a measure of coincidence (probability: 50\%).
Since only frames are missing from the stream, it is expected that this type of corruption is not detectable using drops inside \ac{PSFP} as anomaly indicators.
The results in Table \ref{tab:drop_detection_benchmark_results} confirm the expectation.
The recall value is zero for all traffic patterns.

\paragraph{Injection}
Frames are injected into each stream in a uniformly distributed interval.
The distribution length corresponds to the \acl{MTU} of Ethernet ($\approx$\SI{125}{\micro\second} on \SI{100}{\mega\bit\per\second} links).
This is done to increase the number of distinct network states represented in the simulation runs.
Therefore, the size of each frame is also determined by a uniformly distributed payload size ranging between the allowed minimum (\SI{0}{\byte}) and maximum (\SI{1500}{\byte}).
The results in Table \ref{tab:drop_detection_benchmark_results} show that for traffic shaped by \ac{TSN} on link layer (shaped data stream, timed control traffic) the detection of application layer corruptions is worse than of link layer corruptions.
The reason is that the scheduling and shaping of \ac{TSN} enforces a determined stream behavior and thus the influence of application layer corruptions on stream behavior is less powerful.
Scheduling and shaping generate a defined stream behavior and \ac{PSFP} ensures that aspects of the behavior are adhered to.
Timed control traffic is forwarded on pre-defined time windows.
\ac{PSFP} also only allows reception in those time windows shifted by the transmission delay.
All frames injected in application layer will accumulate together with valid frames in the queues of the corruption source.
Injected frames that do not influence transmission delay through differing size will not be dropped in \ac{PSFP}.
In case of the shaped data stream the \ac{CBS} is shaping the traffic exactly on specification.
The \ac{CBM} in \ac{PSFP} is not dropping any frames and the \ac{NADS} has a recall value of 0.
For a stream with strict priority shaping and a fixed frame size (prioritized CAN tunnel) the detection performance is independent of the injection layer.
All frames with a size not matching the \ac{PSFP} configuration are dropped.

\paragraph{Manipulation}
A frame manipulation in the respective stream takes place with a factor of randomness (probability: 50\%).
The manipulation modifies the payload.
The new Ethernet payload has a uniformly distributed size ranging between the minimum (\SI{0}{\byte}) and maximum (\SI{1500}{\byte}).
In some cases, this is the same length as the valid frame, but in most cases the new frame length differs.
Again, no application layer corruptions are detected for the shaped data stream.
\ac{CBS} shapes the frames correctly according to their size.
Differing frame sizes do not change the forwarded bandwidth of the stream in their respective measurement intervals.
Manipulations on timed control traffic and prioritized \ac{CAN} tunnel frames were detected with a high recall value (\cf Table \ref{tab:drop_detection_benchmark_results}).

\paragraph{Reordering}
With coincidence (probability: 50\%) a frame is taken from a stream and then delayed to be inserted after the next frame of the same stream.
Since reordering has no effect on frame size or bandwidth consumption, it is not detectable in most cases without further inspection of, for example, sequence numbers.
For reordering of timed control traffic on application layer, the \ac{TDMA} timing is enforced by the source host on link layer.
In this case, only the loss of the taken frame is an observable misbehavior.
On the other hand, the timing of frames for link layer reordering is not dependent on the configured shaping.
Therefore, those frames miss the period where the respective gate is open in \ac{PSFP} of the switch.
This causes almost all reordered frames of timed control traffic to be dropped.

\paragraph{Rescheduling}
Frames of the corrupted stream get delayed randomly (probability: 50\%) by a uniformly distributed time.
Again, distribution length corresponds to the \acl{MTU} of Ethernet ($\approx$\SI{125}{\micro\second}).
Timed control traffic is the only one with a \ac{PSFP} configuration dependent on timing.
Once again, the source host gates enforce the designed timing for send windows.
Therefore, the wrong timing is not detected for application layer but for link layer corruptions.

\paragraph{Enhancement using TDMA loss detection}
In addition to frame drops, other statistics can also be used as anomaly indicators.
These can be selected in such a way that further modifications become recognizable and the recall increases.
 Further indicators shall be selected  such that  100\% precision is preserved and no false alarm occur.
In the following example, a loss recognition is implemented using a counter for frames that pass through an opened gate. 
This counter exactly increases by one for each opened window with timed control traffic.
If the counter is not increased during an open gate window, an expected packet has not arrived. 
 10 simulations without corruption confirm that this anomaly indicator also attains an FP value of 0 for the static timed control traffic.
Table \ref{tab:tdma_combi_detection_benchmark_results} shows the simulation results for all modifications on timed control traffic with a \ac{NADS} using frame drops and the \ac{TDMA} loss detection in combination.

With the addition of loss detection all 5 link layer corruptions are detectable for critical \ac{TDMA}-based traffic.
The recall minimum is 0.78 and goes up to the maximum of 1.
At the same time, a precision value of 1 highlights a strong and reliable performance for link layer network anomaly detection.
Corruptions with a high recall value indicate a strong deviation from the specified stream behavior and thus can have a strong impact on network performance and competing real-time streams.
The detection performance increases analogous to those higher risk modifications.

Tailored indicators and strict \ac{PSFP} configurations lead to further improvements in the recall. 
In this way, a \ac{TSN} traffic configuration not only minimizes latency and jitter but can also improve security for data streams on the link layer by improving  detectability of and resilience against misbehavior.

\subsection{Discussion}
\label{subsec:detection_discussion}
The results show that anomaly detection without false positives is possible for all five application and link layer corruptions.
On the other hand, our approach is not geared towards minimizing false negatives.
The measured recall values resulting from the number of false negatives show which scenarios could be detected.
The detection performance is influenced by the traffic shaping and the configuration of the \ac{PSFP}.
Application layer corruptions suffer higher false negatives because the traffic shaping is counteracting the misbehavior.
In general, injection and manipulation scenarios are detected with a higher recall value than elimination, reordering, and rescheduling.
Elimination is impossible to detect with a packet drop statistic.
With strict \ac{PSFP} configurations, further statistics can be used as false positive free anomaly detectors and reduce the number of false negatives to expand detectable scenarios.

\section{Macro-Benchmarking a Realistic Vehicle}
\label{sec:evaluation}
We now proceed to evaluating our approach in a realistic environment.
We use the simulation techniques presented in Section \ref{subsec:detection_environment} with a zonal \ac{IVN} topology based on a real car communication matrix.
First, we simulate regular traffic with a strict \ac{PSFP} schedule to confirm the absence of false positives.
Next, we replace legitimate streams from the car network by selected attacks derived from the CIC-IDS~2017~\cite{shg-gniJR-18} dataset to investigate the impact on \ac{PSFP} drops.
CIC-IDS-2017 contains traces from common Internet attacks.
From these, we select attacks which are likely to be seen at connected cars. 
Datasets with attacks specific to \ac{TSN} link layers have not been recorded, yet, and thus could not be fed into our vehicular environment.
All macro-benchmarking simulations with complete configurations and the used datasets are available online\footnote{\url{https://github.com/CoRE-RG/SignalsAndGateways/tree/paper/network_anomaly_detection_in_cars}} for further research.

\subsection{In-Vehicle Network Topology}
\begin{figure}[t]
  \centering
  \includegraphics[width=1\columnwidth, trim= 0.8cm 0.5cm 0.8cm 0.7cm, clip=true]{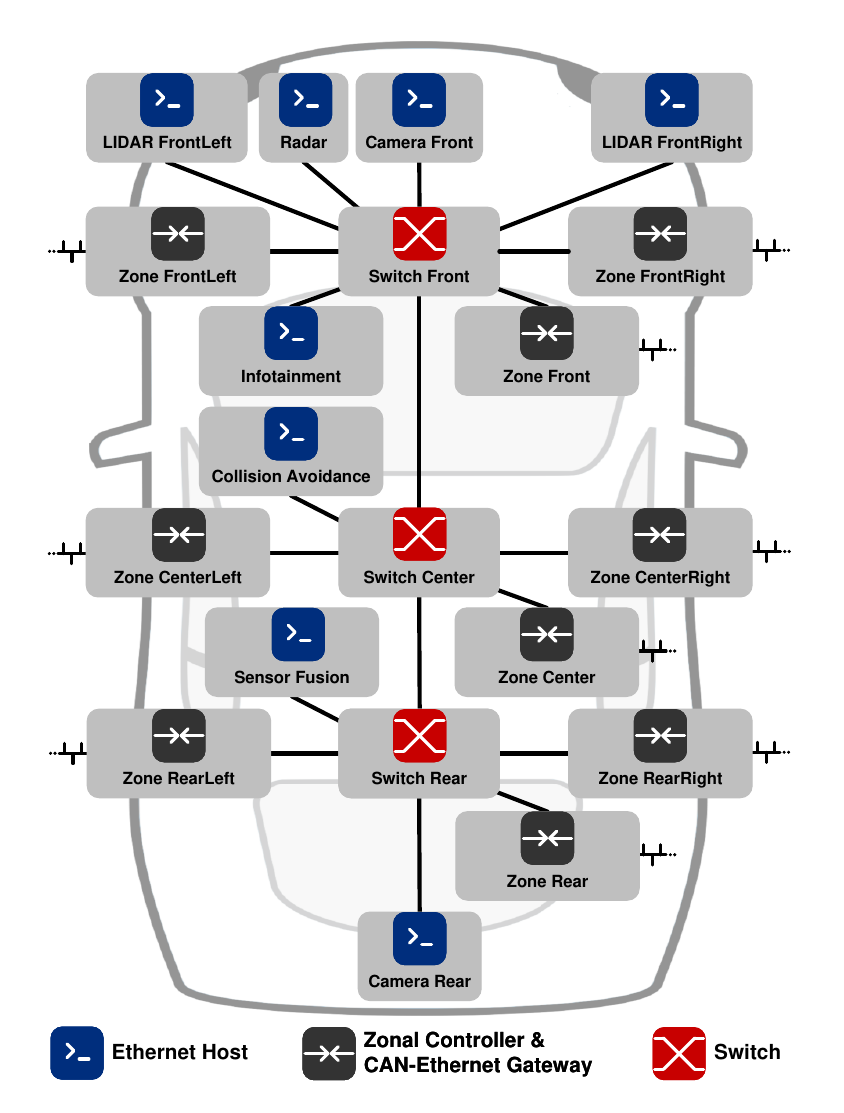}
  \caption{Zonal \ac{IVN} topology based on a real world communication matrix}%
  \label{fig:topology}
\end{figure}

The communication systems of future cars are expected to transition to zonal topologies that still contain legacy devices, before converging to a flat Ethernet backbone~\cite{brkw-aeajr-17}.
Therefore, the network topology of our sample car is split into nine zones (3x~Front, 3x~Center, 3x~Rear) and contains four types of devices (\ac{CAN} hosts, Ethernet hosts, switches, zonal controllers) as visualized in Fig.~\ref{fig:topology}. 
Each zonal controller represents a zone of the topology.
\ac{CAN} hosts are connected via buses to the closest zonal controller depending on physical placement.
The \ac{CAN} traffic is derived from a communication matrix of a production car.
\ac{CAN} messages sent between zones traverse the Ethernet backbone
 (\SI{100}{\mega\bit\per\second} links), which consists of three linked switches.
In addition to the zonal controllers, cameras, LIDARs, a radar, infotainment, collision avoidance, and sensor fusion are also using this backbone for communication.

Future Ethernet backbones contain traffic with heterogeneous \ac{QoS} demands.
Synchronous and asynchronous traffic shaping with priorities will be used to meet these demands~\cite{ieee8021dg-21}.
Therefore, there are three archetypal traffic types in this network:
\begin{itemize}
  \item \textbf{Timed control traffic:}
  Synchronous frames with a size of \SI{64}{\byte} and 802.1Q priority 7 are sent with a period of \SI{1}{\milli\second} from ``Radar'' and ``Sensor Fusion'' to ``Collision Avoidance''.
  A gate control list with the same period implements a \ac{TDMA} schedule in each switch to enable exclusive time slots for this synchronous traffic where the gates of all other priorities are closed.
  % A static \ac{TDMA} schedule in each network device implements this synchronous traffic by configuration of exclusive send time slots.
  \item \textbf{Shaped data streams:}
  Cameras (burst of \SI{27000}{\byte} every \SI{0.033}{\second} per device) and LIDARs (burst of \SI{10500}{\byte} every \SI{0.04}{\second} per device) stream data with 802.1Q priority 5 to the ``Sensor Fusion'' \ac{ECU}.
  For this traffic, class A bandwidth is reserved (\SI{14.144}{\mega\bit\per\second} for each camera stream and \SI{4.544}{\mega\bit\per\second} for each LIDAR stream) on each device along its paths, and a \ac{CBS} shapes the egress.
  \item \textbf{Prioritized \ac{CAN} tunnel:}
  \ac{CAN} messages that are exchanged between devices located in different zones are tunneled via Ethernet.
  There is a total of 416 different \ac{CAN} IDs generated by the \ac{CAN} hosts of which 201 traverse the backbone.
  Their periodicity (\qtyrange{0.01}{2}{\second}) and payload size (\qtyrange{4}{8}{\byte}) is derived from the real car communication matrix.
  Each \ac{CAN} message traversing the backbone is encapsulated in a single Ethernet frame which is padded to a minimum sized frame of \SI{64}{\byte}.
  The frames are mapped to four different 802.1Q priorities (0,1,3,6) depending on their criticality.
\end{itemize}

Each port is  \ac{PSFP}-configured according to the incoming traffic types.
This results in a total of 21 \ac{PSFP} instances in the three switches. 
All configurations use parameters of the network design to check for valid behavior.
Matching the traffic, we use four different types of ingress control:
\begin{itemize}
  \item \textbf{Timing:}
  This configuration uses the static \ac{TDMA} schedule of the \textit{time control traffic} to accordingly set the gate control list that drives the state of the stream gates.
  \item \textbf{Bandwidth:}
  For the \textit{shaped data streams} a metering is configured to enforce a max. incoming bandwidth~\cite{mhks-dpcbm-19}.
  \item \textbf{Frame size:}
  Because a \textit{prioritized CAN tunneling} frame contains only one \ac{CAN} message the size of these Ethernet frames is always \SI{64}{\byte}.
  For the ingress control of this traffic a meter drops all frames larger than \SI{64}{\byte}.
  \item \textbf{Undefined traffic:}
  Ingress control drops traffic without a matching stream filter.
\end{itemize}

\subsection{In-Vehicle Network Scenarios}
In the following, we simulate four scenarios.
All simulations share an identical \ac{TSN} configuration and cover a period of \SI{10}{\second}, which corresponds to the length of the attack sequences.

The first baseline scenario runs with only regular \ac{IVN} traffic to verify the absence of false positives.
Ten simulations with different pseudo random number generator seeds are executed to cover different cases.
In all simulations without corruption, not a single frame is dropped by any \ac{PSFP} instances (\cf Table \ref{tab:case_study_regular_traffic}).
This consistently indicates zero \ac{FP} and full precision.

\begin{table}[h]
  \caption{Sum of \ac{PSFP} frame drops (FPs) in the baseline scenario}
  \label{tab:case_study_regular_traffic}
  \centering
  \begin{tabularx}{\columnwidth}{lYYY}
  \toprule
  \multirow{2}*{\textbf{Simulation run}} & \multicolumn{3}{c}{\textbf{Sum of frames in all \ac{PSFP} instances}}\\
  \cmidrule(rl){2-4}
                                         & \centering{\textbf{Arrived}} & \textbf{Forwarded} & \textbf{Dropped (\ac{FP})}\\
  \midrule
  \#0 --- \#9 & \num{3166150} & \num{3166150} & \num{0} \\
  \bottomrule
  \end{tabularx}
\end{table}

\begin{figure*}
  \centering
  \begin{tikzpicture}
      \begin{groupplot}[
      group style={
        group size=3 by 4, 
        horizontal sep=12pt,
        vertical sep=12pt,
        xticklabels at=edge bottom,
        yticklabels at=edge left,
      },
      width=0.383\linewidth,
      height=80pt,
      change x base,
      xtick distance={1},
      xtick pos=bottom,
      xmin=-0.5,
      xmax=10,
      change y base,
      y SI prefix=kilo,
      ytick distance={1000},
      ymin=0,
      ymax=1000,
      groupplot xlabel={Simulation time [\si{\second}]},
      groupplot ylabel={Frame size [\si{\kilo\byte}]},
      group/only outer labels,
      legend image code/.code={\draw [#1] (0cm,-0.1cm) rectangle (0.1cm,0.2cm); },
      ]
    \nextgroupplot[align=center, title={SSH-Patator\\Section \ref{subsubsec:attack_sshpatator}}]%0
      \node[coredarkgray] at (axis cs:5,800) {Attack trace};
      \addplot[ycomb,
        coreblue,
        ultra thick
      ] table [x, y, col sep=comma] {data/SSHPatator.csv};
      \coordinate (top) at (rel axis cs:0,1);
    \nextgroupplot[align=center, title={Web Attack Brute Force\\Section \ref{subsubsec:attack_webattackbruteforce}}]%0
      \node[coredarkgray] at (axis cs:5,800) {Attack trace};
      \addplot[ycomb,
        coreblue,
        ultra thick
      ] table [x, y, col sep=comma] {data/WebAttackBruteForce.csv};
    \nextgroupplot[align=center, title={DoS Slowloris\\Section \ref{subsubsec:attack_dosslowloris}}]%0
      \node[coredarkgray] at (axis cs:5,800) {Attack trace};
      \addplot[ycomb,
        coreblue,
        ultra thick
      ] table [x, y, col sep=comma] {data/DoSSlowloris.csv};
    \nextgroupplot%1
      \node[coredarkgray] at (axis cs:5,800) {Radar $\rightarrow$ Collision Avoidance};
      \node[coredarkgray] at (axis cs:5,500) {Timed Control Traffic};
      \addplot[ycomb,
        corered,
        ultra thick
      ] table [x, y, col sep=comma] {data/SSHPatator_Radar_manipulation.csv};     
    \nextgroupplot%1
      \node[coredarkgray] at (axis cs:5,800) {Radar $\rightarrow$ Collision Avoidance};
      \node[coredarkgray] at (axis cs:5,500) {Timed Control Traffic};
      \addplot[ycomb,
        corered,
        ultra thick
      ] table [x, y, col sep=comma] {data/WebAttackBruteForce_Radar_manipulation.csv};
    \nextgroupplot%1
      \node[coredarkgray] at (axis cs:5,800) {Radar $\rightarrow$ Collision Avoidance};
      \node[coredarkgray] at (axis cs:5,500) {Timed Control Traffic};
      \addplot[ycomb,
        corered,
        ultra thick
      ] table [x, y, col sep=comma] {data/DoSSlowloris_Radar_manipulation.csv};
    \nextgroupplot%2
      \node[coredarkgray] at (axis cs:5,800) {Camera Front $\rightarrow$ Sensor Fusion};
      \node[coredarkgray] at (axis cs:5,500) {Shaped Data Stream};
      \addplot[ycomb,
        green,
        ultra thick
      ] table [x, y, col sep=comma] {data/SSHPatator_CamFront_manipulation_switch_center.csv};
    \nextgroupplot%2
      \node[coredarkgray] at (axis cs:5,800) {Camera Front $\rightarrow$ Sensor Fusion};
      \node[coredarkgray] at (axis cs:5,500) {Shaped Data Stream};  
    \nextgroupplot%2
      \node[coredarkgray] at (axis cs:5,800) {Camera Front $\rightarrow$ Sensor Fusion};
      \node[coredarkgray] at (axis cs:5,500) {Shaped Data Stream}; 
    \nextgroupplot%3
      \node[coredarkgray] at (axis cs:5,800) {Gateway FrontLeft $\rightarrow$ Gateway *};
      \node[coredarkgray] at (axis cs:5,500) {Prioritized \ac{CAN} Tunnel};
      \addplot[ycomb,
        corered,
        ultra thick,
      ] table [x, y, col sep=comma] {data/SSHPatator_GwFrontLeft_injection.csv};
    \nextgroupplot%3
      \node[coredarkgray] at (axis cs:5,800) {Gateway FrontLeft $\rightarrow$ Gateway *};
      \node[coredarkgray] at (axis cs:5,500) {Prioritized \ac{CAN} Tunnel};
      \addplot[ycomb,
        corered,
        ultra thick,
      ] table [x, y, col sep=comma] {data/WebAttackBruteForce_GwFrontLeft_injection.csv};
    \nextgroupplot%3
      \node[coredarkgray] at (axis cs:5,800) {Gateway FrontLeft $\rightarrow$ Gateway *};
      \node[coredarkgray] at (axis cs:5,500) {Prioritized \ac{CAN} Tunnel};
      \addplot[ycomb,
        corered,
        ultra thick,
        forget plot
      ] table [x, y, col sep=comma] {data/DoSSlowloris_GwFrontLeft_injection.csv};
    \coordinate (bot) at (rel axis cs:1,0);
    \addlegendimage{coreblue, fill=coreblue, ultra thick, -}
    \label{plots:trace}
    \addlegendimage{corered, fill=corered , ultra thick, -}
    \label{plots:drops}
    \addlegendimage{green, fill=green, ultra thick, -}
    \label{plots:drops_c}
  \end{groupplot}
  \path (top|-current bounding box.north)--coordinate(legendpos)(bot|-current bounding box.north);
  \matrix[
      matrix of nodes,
      anchor=south,
      draw,
      inner sep=0.15em,
      draw
    ]at([xshift=0ex,yshift=-60ex]legendpos)
    {
      \ref{plots:trace}& Attack frame transmission &[5pt]
      \ref{plots:drops}& Switch Front frame drop &[5pt]
      \ref{plots:drops_c}& Switch Center frame drop &[5pt]
      \\};
  \end{tikzpicture}
  \caption{
    Selected attacks (SSH-Patator, Web Attack Brute Force, DoS Slowloris) from the CIC IDS 2017 dataset abuse different traffic types (timed control traffic, shaped data stream, prioritized CAN tunnel) of three \acp{ECU} (Radar, Camera Front, Gateway FrontLeft) and their impact on \ac{PSFP} frame drops in Switch Front and Switch Center
  }
  \label{fig:cicids2017selection}
\end{figure*}
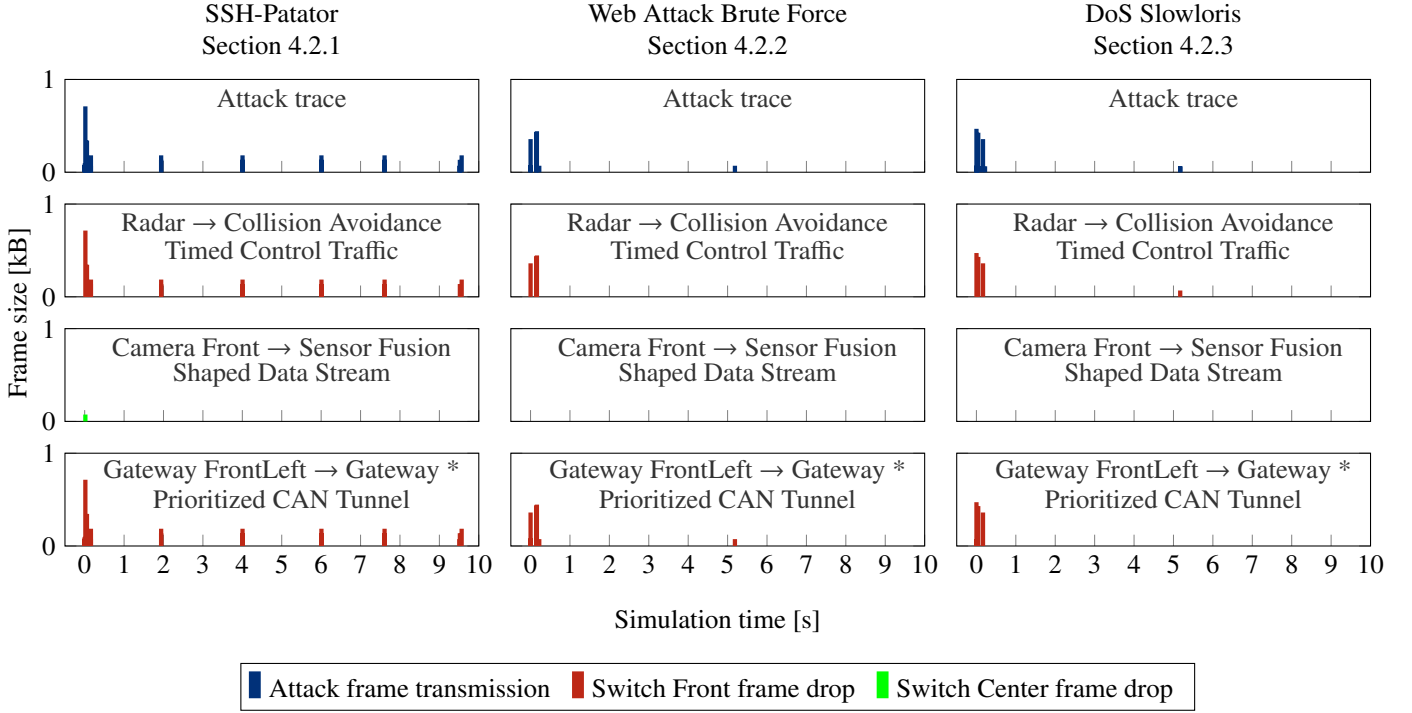

The next three scenarios use attack samples exported from CIC-IDS~2017~\cite{shg-gniJR-18} (SSH-Patator, Web Attack Brute Force, DoS Slowloris).
All attacks included in CIC-IDS~2017 are application layer corruptions from the \ac{PSFP} perspective.
Trace excerpts of these attacks contain frames targeting the same destination.
This enables the matching into \ac{IVN} streams by replacing MAC addresses and adding a Q-Tag (VLAN and priority).
The three attack traces contain 8 to 26 frames.
In each simulation, one of the three participant (Radar, Camera Front, Zonal Controller FrontLeft) gets corrupted and performs an attack.
Each scenario contains three simulation runs, in which one network participant is the source of a modified traffic type (timed control traffic, shaped data stream, prioritized \ac{CAN} tunnel).

\subsubsection{Attacks using SSH-Patator}
\label{subsubsec:attack_sshpatator}
Patator is a tool for brute force attacks using a multitude of protocols (e.g., FTP, SSH, DNS). 
The CIC-IDS 2017 trace excerpt used in this scenario contains traffic of a SSH login brute force attack.

The sources of corrupted streams  during three different simulations are Radar, Camera Front and Zonal Controller FrontLeft.
Switch Front is the first hop for all corrupted streams and thus the first point where \ac{PSFP} is applied.
The first column of Fig.~\ref{fig:cicids2017selection} shows the SSH-Patator attack trace content and the respective \ac{PSFP} drops in Switch Front that were collected during the three different simulations.

\paragraph*{Radar $\rightarrow$ Collision Avoidance (Control traffic)}
The regularly timed control traffic from Radar to Collision Avoidance is replaced by SSH-Patator traffic.
The send timing in the source Radar is ensured through the \ac{TSN} scheduling and shaping configuration.
Nevertheless, all frames miss their arrival time window on Switch Front because their size is larger than the regular \SI{64}{\byte}.
Therefore, the gates in the responsible \ac{PSFP} instance are closed when the frames arrive, and they get dropped.
The \ac{NADS} detects the stream corruption. 

\paragraph*{Camera Front $\rightarrow$ Sensor Fusion (Data stream)}
SSH-Patator traffic replaces the regularly shaped data stream.
Not a single frame is dropped in Switch Front, so a detection is not possible.
The reason is the same as in the application layer corruption micro-benchmarks (s. Section \ref{sec:benchmark_v2}):
The \ac{CBS} at Camera Front shapes the attack traffic exactly according to the specification.

In this case, however, one frame of this stream gets dropped in \ac{PSFP} of Switch Center, which is highlighted in green.
Interference with other traffic between Switch Front and Switch Center causes the \ac{CBS} in Switch Front to transmit a burst that is not allowed by the \ac{PSFP} configuration in Switch Center.
Since \ac{PSFP} operates on individual streams, it is still possible to infer the misbehaving stream.
Following this direction can improve the detection of bogus shaped data streams.

\paragraph*{Gateway FrontLeft $\rightarrow$ Gateway * (\ac{CAN} tunnel)}
One \ac{CAN} tunnel stream is replaced by the SSH-Patator traffic.
Here the regular traffic again consists of minimal Ethernet frames (\SI{64}{\byte}) each containing only one CAN message at a time.
The \ac{PSFP} instance drops larger frames because the flow meter is configured to enforce the minimum Ethernet frame size.
All attack frames are larger and get dropped in Switch Front.
The stream corruption is detected.

\subsubsection{Brute Force Web Attacks}
\label{subsubsec:attack_webattackbruteforce}
This trace excerpt contains an attempt to login via HTTP using a password list.
Again, the second column in Fig.~\ref{fig:cicids2017selection} shows the Web  Brute Force attack trace and the respective \ac{PSFP} drops in Switch Front that were collected during three simulations.

\paragraph*{Radar $\rightarrow$ Collision Avoidance (Control traffic)}
The timed control traffic of Radar is replaced by traffic from a Web  Brute Force attack trace.
The larger frames are dropped again because they miss the time window for allowed reception in Switch Front.
The frames at \SI{0.2}{\second} and \SI{5.2}{\second} simulation time, however, are not dropped.
They are small enough (\SI{74}{\byte}) to meet the configured timing requirement and arrive just before the responsible \ac{PSFP} gate closes.
Despite of this, the corruption is detectable because some attack frames are dropped.

\paragraph*{Camera Front $\rightarrow$ Sensor Fusion (Data stream)}
The shaped data stream of Camera Front is corrupted by Web Attack Brute Force traffic.
As before, the Camera Front \ac{CBS} shapes the attack traffic exactly according to the specification.
Hence, Switch Front \ac{PSFP} drops, and no severe misbehavior occurs via concurrency with other streams after Switch Front. Therefor, no frame is deleted at any location and detection remains unsuccessful.

\paragraph*{Gateway FrontLeft $\rightarrow$ Gateway * (\ac{CAN} tunnel)}
Web  Brute Force attack traffic is injected into a \ac{CAN} tunnel stream.
All attack frames, even small ones (at \SI{0.2}{\second} and \SI{5.2}{\second}), are dropped because they are larger (\SI{74}{\byte}) than Switch Front \ac{PSFP} allows (\SI{64}{\byte}).
The system fully detects the stream corruption.

\subsubsection{Attacks using DoS Slowloris}
\label{subsubsec:attack_dosslowloris}
This trace contains an attack that tries to stress a server with multiple simultaneous HTTP connections.
The third column in Fig.~\ref{fig:cicids2017selection} shows the DoS Slowloris attack trace  and the respective \ac{PSFP} drops in Switch Front that were collected during three simulations.

\paragraph*{Radar $\rightarrow$ Collision Avoidance (Control traffic)}
DoS Slowloris traffic replaces the regularly timed control traffic from Radar.
All larger frames are dropped.
The trace contains minimal Ethernet frames (\SI{64}{\byte}), which are not dropped by \ac{PSFP}.
An exception in this case are two minimal frames at \SI{5.17}{\second}
 sent by the source app with a gap of \SI{229}{\micro\second}. Both frames arrive in the outgoing queue of Radar before the send interval starts.
After the gate for the outgoing queue opens, both frames are transmitted consecutively.
The first frame passes through the Switch Front \ac{PSFP}.
The second frame gets dropped on arrival because it no longer fits into the Switch Front \ac{PSFP} time window.
Corruption detection is successful.

\paragraph*{Camera Front $\rightarrow$ Sensor Fusion (Data stream)}
The shaped data stream corrupted by DoS Slowloris shows the same results as for SSH-Patator and Web Attack Brute Force.
No detection is possible because no frames are dropped.

\paragraph*{Gateway FrontLeft $\rightarrow$ Gateway * (\ac{CAN} tunnel)}
One prioritized \ac{CAN} tunnel stream is replaced by DoS Slowloris.
All frames that are not minimal (\SI{64}{\byte}) get dropped.
Both minimum frames at \SI{5.17}{\second} can pass the Switch Front \ac{PSFP}, because in this case only the frame size is enforced.
Nevertheless, the attack is detected.

\subsection{Findings}
The macro-benchmarks confirm the results previously found in the application layer micro-benchmarks (s. Section \ref{sec:benchmark_v2}) for all evaluated scenarios.
We find zero false positives in a realistic automotive setup with a strict, consistent \ac{PSFP} configuration based on a real car communication matrix.
Real attacks are reliably detected in all scenarios, except for shaped data stream corruptions. Most importantly, traffic exceeding specification limits such as volumetric DoS attacks is fully identified. 
The data stream corruptions are not detectable on the link layer because the CBS traffic shaping at the attack source nodes enforce valid bandwidth usage.
Undetected corruptions, however, are not breaking with configured specifications and should have lower risk in damaging \ac{QoS} of concurrent streams.
Additionally, one scenario showed a successful detection of a corrupted shaped data stream through interference with concurrent traffic on the path to the second switch.
This opens directions for further improving the detection in the future.

\section{Security for Cars and Related Work}
\label{sec:background_v2_security}

The seminal work by Checkoway~\etal \cite{cmkas-ceaas-11} examines interfaces that are part of the attack surface of a car.
These interfaces are classified into three categories: Physical access (ODB-II, CD, USB), short distance wireless access (Bluetooth, Wi-Fi, Remote-Keyless-Entry) and long-distance wireless access (GPS, digital radio, mobile services).
The authors could gain access to the on-board network in each category using reverse engineering and debugging.
Hence, it is safe to expect that unauthorized access is possible to all in-car components.
Each communication source within the vehicle could become corrupted and send irregular traffic into the \ac{IVN}.

Miller and Valasek~\cite{mv-sraas-14, mv-reupv-15} show that modern cars are vulnerable to attacks.
In their work, they describe how they obtain control of an unaltered passenger vehicle.
The entry point is the cellular connection of the car infotainment system.
They use the infotainment to send manipulative messages into the \ac{IVN}.
Those messages enabled remote control of safety-critical functions such as engine and brakes.
When protective measures are overcome, a \acf{NADS} is important to detect such illegitimate messages and report incidents for possible countermeasures.

\subsection{Security in Future Cars}

In addition to \ac{TSN} and \ac{SDN} technologies, several network technologies entered the design discussion for enabling safe and secure communication with high performance for \acp{IVN} in recent years.
The industry standard \ac{SOME/IP}~\cite{autosar-someip-17} introduces \ac{SOA} for \acp{IVN}.
Even containerized services are discussed  for deploying dynamic functions in high performance \acp{ECU}~\cite{kakww-dsosa-19}.

Multi-sided measures are needed to protect future cars against attacks~\cite{m-saejr-17}.
Furthermore, new regulations like the European Cybersecurity Act~\cite{eu-ecaJR-19} and guidelines like ISO/SAE 21434 demand extensive security features such as updates, monitoring, and incident management/response for the entire lifecycle of future vehicles~\cite{unece-wp29,iso-sae-21434}.
To protect the security of future \acp{IVN}, the appropriate tools from the toolbox of established network security mechanisms, such as encryption, authentication, firewalling, and intrusion detection, must be adapted and implemented.~\cite{k-gcnjr-17}.

One of the first steps for securing information systems is the risk assessment~\cite{rp-spcJR-20}.
Monteuuis~\etal \cite{mbzls-ssajr-18} propose a systematic thread analysis and risk assessment framework for autonomous cars.
They build upon traditional methods to support consideration of the safety implications through automotive attacks.
To compute risk values, they use severity, observation, controllability and the attack likelihood.
Further, they clarify the importance of countermeasures for risk assessment: 
``[...] risk analysis is an iterative process that ends once countermeasures have been applied to critical threats until the risk value converges to an acceptable level.''
Countermeasures are dependent on the detection of threads.
Thus, anomaly detection and comprehensive information on observability of corruptions are important for precise risk assessment and vehicle safety and security.
Therefore, we classify link layer effects of anomalies and benchmark the detection performance.

The work of Bernardini~\etal \cite{bac-spvjr-17} inspects the main security and privacy issues of vehicular communication and explores related research.
Authors found that effective network monitoring is one of the open problems for the security of modern cars.
Our work tries to fill this gap by targeting link layer anomaly detection and detailed measurements.

Rumez, Grimm~\etal \cite{rgks-oasjr-20} discuss the security implications of \acp{SOA}.
They show that, in general, the protection of \acp{IVN} must be achieved by integrating different layers of protection.
The review presents approaches on automotive firewalls, \acp{IDS} and \ac{IAM}.
The target of all reviewed \acp{IDS} is the \ac{CAN} bus protocol.
In contrast, we investigate a real-time Ethernet backbone.

Pesé~\etal \cite{psz-hcdjr-17} present a hardware/software co-design for an automotive firewall.
They demonstrated how an embedded design can fulfill automotive requirements in a cost-efficient way.
The investigations on the performance of our \ac{NADS} are simulation-based and do not address specific implementation aspects related to hardware or software resources deployable in a production vehicle.
Nevertheless, the design we propose is efficient on in-car resources because no specialized \ac{NADS} devices are needed for operation.
Instead, we utilize \ac{TSN} mechanisms, already discussed for on-board networks, to additionally detect anomalies.

Firewalls are often a part of gateways between domain- and system borders.
Those gateways also fulfill roles such as proxy, access control~\cite{rdgks-iabac-19}, and intrusion detection~\cite{ymhs-tbijr-19}.
Gateways are also introduced as pure security nodes~\cite{so-sagjr-14}.
They can enforce static specified communication streams between their in- and outputs.
Thus, those gateways are able to prohibit unknown communication.
This inline flow-control can be fulfilled by switches in \ac{SDN}-based Ethernet networks.
In past work, we analyzed the security implications of control flow embedding strategies~\cite{hmks-stsdn-22}.
Now, we show how valid flows can be monitored and misbehaving flows can be detected in \acp{IVN} using \ac{TSN} and \ac{SDN} principles.

Langer~\etal \cite{lss-eacdc-19} show how security of complete vehicle fleets could be managed over lifetime.
The proposed \ac{ACDC} adapts established IT infrastructure technologies to the nomadic nature of vehicle fleets.
Such cloud services can be used to collect reports of severe incidents detected by an \ac{IDS}.
This enables large-scale countermeasures.
Data from our \ac{NADS} approach is a substantial source of information for global security infrastructures such as this \ac{ACDC}.

In previous works, we explored how \ac{SDN} can be integrated in a real-time Ethernet \ac{IVN} and how SDN can enhance its resilience~\cite{rhmks-rapesc-20,hmks-stsdn-22}.
Furthermore, we demonstrated a prototype of an \ac{IVN} that enables protection, monitoring, detection, incidence management, and countermeasures in a real production car~\cite{mhlsd-dsivi-20}.
The implementation uses \ac{SDN}, \ac{AD}, secure gateways, hypervisors, dynamic orchestration of applications and cloud services to enable protection and countermeasures.
This prototype uses a machine learning based \ac{NADS} which is prone to frequent false positive alarms.
In this work, we show a \ac{NADS} that is robust against false positive alarms.

\subsection{Anomaly Detection in Future Cars}

Dibaei~\etal \cite{dzjal-adijr-20} survey attacks and defenses on intelligent connected vehicles.
They categorize security-related attacks into cryptography, network security, software vulnerability detection, and malware detection.
Authors further stress that \acp{IDS} are the most effective countermeasure for network security.
They discuss the advantages and shortcomings of signature-based and anomaly-based IDS approaches,  
in particular a high false negative rate for signature-based \ac{IDS} and a high false positive rate for anomaly-based \ac{IDS}.
In our anomaly-based approach, we use the accurate \acf{PSFP} configuration as the baseline.
Our concept enables link layer anomaly detection without false positives.

There are many anomaly detection algorithms at hand to describe and evaluate a regular state~\cite{bbk-nadms-14}.
They can be classified in statistical (e.g., Signal Processing), classification based (e.g., Support Vector Machine), clustering-based (e.g., k-Means), soft computing (e.g., Neural Network), knowledge-based (e.g., Rule and expert-system), or combination learners (e.g., Hybrid).
All \acp{AD} algorithms try to fingerprint normal behavior.
They are trained or configured to distinguish between normal and abnormal behavior.
One key challenge for anomaly detection is to keep the false-positive rate as low as possible. 
Any unclear or ill-defined distinction or rare event may lead to false positive anomaly reports and degrade detection quality.
By using \ac{TSN} mechanisms which are designed to also configure the behavior of the traffic our approach operates at an intersection of anomaly detection and traffic shaping.
This enables a clear definition of normal real-time traffic behavior and thus zero false positives.
On the other hand, the rate of false negatives is not guaranteed to be leading. 
It is no inherent property of our approach to maximize the number of abnormal behavior that is detectable.
Also, in contrast to other detection algorithms, our approach is naturally limited to \ac{TSN} networks.

Rajbahadur~\etal \cite{rmwh-sadjr-18} dedicate a survey to anomaly detection for connected vehicles.
They discuss a large body of work in the spectrum of anomaly detection for current cars.
Most of the work addressing the use of anomaly detection systems in \acp{IVN} aim at traditional \ac{CAN} bus infrastructures.
Ethernet \acp{IVN}, though, have very different communication patterns leading to different requirements on a \ac{NADS}.
Our approach aims at Ethernet using \ac{PSFP}---an integral part of \ac{TSN}.

Waszecki~\etal \cite{wmslk-sdvtm-17} present a distributed \ac{IVN} traffic monitoring via behavioral analysis on \ac{CAN} buses.
Their decentralized approach enhances reliability of attack detection and reduces implementation costs.
Our approach is also decentralized and cost-effective by using every port on every switch in the network as an anomaly sensor.
No additional network nodes are required to implement the anomaly sensors into an \ac{IVN}.

The work that exploit \ac{CAN} protocol properties for anomaly detection in \acp{IVN} are based on \ac{CAN} ID sequences~\cite{ms-adcJR-17} or remote frames~\cite{ljk-oniJR-17}. This achieves high detection accuracy for message injection attacks.
Our micro benchmarks evaluate the performance also for the orthogonal link layer anomaly classes including elimination, injection, manipulation, reordering, and rescheduling. 

Paul~\etal \cite{pi-annJR-21} use an artificial neural network for \ac{CAN} anomaly detection with high detection performance for \ac{DoS} and Fuzzy attacks.
Islam~\etal \cite{irym-gbiJR-22} present a graph-based intrusion detection system that show high detection performance for \ac{DoS}, fuzzy, replay, and spoofing attacks.
Technically parts of those approaches are transferable to Ethernet based \acp{IVN}, but they do not operate without false alarms.
Our focus is on exploiting \ac{TSN} properties that enable zero \ac{FP} anomaly detection.

Khraisat~\etal \cite{kgvk-sidJR-19} discuss the increasing sophistication of cyber-attacks and urge the need for accurate intrusion detection.
Authors present a taxonomy of contemporary \ac{IDS}, including signature-based and anomaly-based intrusion detection systems, and provide a comprehensive review of recent work, commonly used datasets for evaluation, evasion techniques used by attackers, and future research challenges.
Our work deviates strongly from classical network environments in the application area. 
The behavior of attackers on \acp{IVN} with \ac{TSN} is still unspecified and datasets must additionally represent the regular real-time Ethernet  behavior.

 Ring~\etal \cite{rwslh-snbjr-19} give an overview of network-based data sets for testing and evaluating \acp{IDS}.
They analyze which properties the individual data sets fulfill.
To the best of our knowledge, there are no datasets that contain attacks  specific to real-time Ethernet and its link layer \ac{QoS}, which is a major attack surface in cyber physical systems such as cars.
In our macro-benchmarks, we integrate traffic of the CIC-IDS~2017~\cite{shg-gniJR-18} dataset, which includes the most common attacks according to the 2016 McAfee report.

\section{Conclusions and Outlook}
\label{sec:conclusion}
Communication flows for vehicle control are pre-defined to a large extend, and \ac{TSN} \acf{PSFP} builds its traffic control rules for all incoming traffic streams on this knowledge.
In this paper, we showed how the \ac{PSFP} control rules can serve as the basis for a \acf{NADS}.
The \ac{PSFP} configuration serves as an implicit distinction between normal and malicious network behavior on the data link layer.

 Conceptually and by simulation experiments we could show that this anomaly detection remains free of false positives, provided the \ac{PSFP} indicators are defined concisely and correctly.
In all baseline simulations, not a single valid frame was dropped by any \ac{PSFP} instance.

The efficiency to detect misbehavior depends on the origin of frame corruption, its link layer anomaly type, and its traffic type.
Overall, corruptions on the link layer are more reliably detected than application-specific misbehavior.
Most of the critical impairments in real-time networks impact the link layer, since it accumulates the risk of breaking \ac{QoS} guarantees for concurrent streams. In particular, this holds for any kind of volumetric DoS attacks.
In addition, there are \ac{TSN} traffic shaping mechanisms that enable a detection of application layer corruptions by frame drop indicators.
We showed by example that such indicators, which also remain free of false positives, can be successfully deployed and improve the detection rate.

We also presented macro-benchmarks, in which real attack traces were played back in a simulated zonal \ac{IVN} topology based on a real-world in-car communication matrix.
Our results indicate that the \ac{NADS} works well in realistic \ac{IVN} environments and consistently avoids false positives. Our simulations also revealed the limits of our approach and characterized the bogus traffic classes which remain undetectable.

Our work opens three directions for future research.
\begin{enumerate}
\item Options for improving detection performance should be explored. 
Additional \ac{PSFP} anomaly indicators are promising candidates to improve the accuracy without introducing false positives.
This  includes the investigation of further \ac{TSN} forwarding procedures such as Asynchronous Traffic Shaping and  
the effects of chained \ac{PSFP} instances in switches on stream paths.

\item Evaluation in more complex, realistic scenarios are likewise desirable. 
We will investigate how complete \ac{IDS} datasets can be used for comprehensive evaluations on detection performance in the context of \ac{TSN}-based \acp{IVN}.
This may include the creation of a special attack dataset for \acp{IVN}~\cite{mhlks-fsaad-24}.
In addition, the proposed \ac{NADS} mechanism should be examined in a \ac{TSN} testbed to validate its effectiveness and measure resource consumption.

\item An integration with orthogonal \ac{NADS} concepts external to the core network, which use extrinsic mechanisms such as machine learning to define regular pattern, should be evaluated for extending the intrusion detection system to application layer characteristics.
\end{enumerate}

\section*{Acknowledgments}
This work was funded in parts by the German Federal Ministry of Education and Research (BMBF) within the SecVI project (grant no. 16KIS0815K).

%%%% 	BibTeX		%%%%
\bibliographystyle{elsarticle-num}
\bibliography{HTML-Export/all_generated,rfcs,bib/special,local}

\end{document}